\documentclass{npcs}
\usepackage{graphics} %   <------- for LatTeX + PS
\usepackage{epsfig}
\usepackage{axodraw}
\usepackage{multicol}
\usepackage{amsmath}
\begin{document}

\runauthor{T.V. Shishkina, V.A. Mossolov, I.B. Marfin}

\runtitle{Anomalous quartic boson couplings via
$\gamma\gamma\rightarrow W^+W^-$ and $\gamma\gamma\rightarrow W^+W^-Z$
at the TESLA kinematics}

\begin{topmatter}
\title{ Anomalous quartic boson couplings via
$\gamma\gamma\rightarrow W^+W^-$ and $\gamma\gamma\rightarrow W^+W^-Z$
at the TESLA kinematics}

\author{Author I.B. Marfin}
\institution{NCHEP}
\address{153 Bogdanovitcha str.,220040 Minsk, Belarus}
\email{marfin@hep.by}

\author{Author V.A. Mossolov}
\institution{NCHEP}
\address{153 Bogdanovitcha str.,220040 Minsk, Belarus}
\email{mos@hep.by}

\author{Author T.V. Shishkina}
\institution{NCHEP}
\address{153 Bogdanovitcha str.,220040 Minsk, Belarus}
\email{shishkina@hep.by}

\begin{abstract}
The production of two and three electroweak gauge bosons in the
high-energy $\gamma\gamma$ collisions
gives the well opportunity to probe anomalous quartic gauge boson couplings.
The influence
of five possible anomalous couplings on the cross sections for
$W^+W^-$, $W^+W^-\gamma$, $W^+W^-Z$ productions has been investigated
at the TESLA kinematics ($\sqrt{S}\sim 1$ TeV).
There are the reasonable
discriminations between various anomalous
contributions.

%In considering  $\gamma\gamma\rightarrow W^+W^-Z$ one
%has to take into account radiative corrections to
%$\gamma\gamma\rightarrow W^+W^-$. The last are very considerable at
%high energies and greatly contribute to the differential cross
%section $d\sigma_{\lambda_1\lambda_2\lambda_3\lambda_4}$ at various
%polarizations of initial photons and final bosons.

\end{abstract}
\end{topmatter}

\newpage

\section{Introduction}

$ $

The multiple vector-boson production will be a
crucial test of the gauge structure of the Standard Model since the
triple and quartic vector-boson couplings involved in this kind of
reaction are constrained by the $SU(2)\bigotimes U(1)$ gauge
invariance. The production of several vector bosons is the best place to
search directly for any anomalous behavior of triple and quartic
couplings.

Any small deviation from the Standard Model predictions for these
couplings spoils the cancellations of the high energy
behaviour between the various diagrams, giving rise to an
anomalous growth of the cross section with energy \cite{c5}. It is
important to measure the vector-boson selfcouplings and search
deviations from the Standard Model (SM).

The trilinear and quartic gauge boson couplings affect different
aspects of the electroweak interactions. The trilinear couplings
directly test devations from the non-Abelian structures of
the SM \cite{c4}. On the contrary, the quartic gauge boson
couplings leads to direct  electroweak symmetry breaking,
in  particular to  the scalar  sector of  the theory  or, more
generally, to new physics of electroweak gauge bosons.

Thus there is the possibility that only the quartic couplings have
values deviated from their SM values while the
trilinear couplings don't anomalous affect on the $W^+W^-$ and $W^+W^-Z$
$(W^+W^-\gamma)$ productions \cite{c6}. Since the
mechanism of symmetry breaking isn't revealed completely so
anomalous quartic
gauge bosons can explain it and provide the first evidence of
"new physics" in this sector of
the electroweak theory.

Using the future high-energy linear $e^+e^-$-collider, one can
obtain the colliding $\gamma e$ and $\gamma\gamma$ beams with
almost  the same energies as in $e^+e^-$-
collisions and with high luminosity \cite{c5}.
Processes $\gamma\gamma\rightarrow W^+W^-$ and
$\gamma\gamma\rightarrow W^+W^-Z$  at high energies
will give unique possibility of quartic couplings investigation due to
relatively large cross sections and low background for
$WW$ and $WWZ$  productions. In this paper boson productions
will be considered at energies $\sqrt{S}$ about
$1$ TeV corresponding to the TESLA kinematics.

In the following section we review the various types
of anomalous quartic couplings that might be expected
in extensions of the SM. In the section $3$ the
numerical results illustrating the effect of the
anomalous couplings on the $WW$ and $WWZ$ cross sections
as well as conclusions are presented.

\section{Feynman rules for anomalous quartic gauge boson couplings}

$ $

In order to construct the structures contained anomalous
quartic gauge boson couplings where one photon is involved at
least one has to consider the operators with the the
lowest dimension of $6$ \cite{c2,c1}. That is required for a
custodial $SU(2)_c$
symmetry to have the $\rho=M_W^2/(M_Z^2\cos^2{\theta_W})$
parameter close to $1$. Thus we consider the $6$-dimensional
operators \cite{c1}
\begin{eqnarray}\label{a1}
\begin{array}{c}
\displaystyle {\cal L}_0 = -\frac{e^2}{16\Lambda^2}a_0F^{\mu\nu}
F_{\mu\nu}\bar{W}^{\alpha}\bar{W}_{\alpha}, \\  \\
\displaystyle {\cal L}_c = -\frac{e^2}{16\Lambda^2}a_cF^{\mu\alpha}
F_{\mu\beta}\bar{W}^{\beta}\bar{W}^{\alpha}, \\  \\
\displaystyle \tilde{\cal L}_0 = -\frac{e^2}{16\Lambda^2}\tilde{a}_0
F^{\mu\alpha}\tilde{F}_{\mu\beta}\bar{W}^{\beta}\bar{W}^{\alpha}, \\  \\
\displaystyle {\cal L}_n = -\frac{e^2}{16\Lambda^2}a_n\epsilon_{ijk}
F^{\mu\nu}W_{\mu\alpha}^iW_{\nu}^jW^{\alpha,\, k }, \\     \\
\displaystyle \tilde{\cal L}_n = -\frac{e^2}{16\Lambda^2}\tilde{a}_n
\epsilon_{ijk}\tilde{F}^{\mu\nu}W_{\mu\alpha}^iW_{\nu}^j
W^{\alpha,\, k },
\end{array}
\end{eqnarray}
where we introduce the triplet of gauge bosons
\begin{eqnarray}\label{a2}
\begin{array}{c}
\displaystyle \bar{W}_{\mu} = \left(\frac{1}{\sqrt{2}}(W^+_{\mu}+
W^-_{\mu}),\frac{i}{\sqrt{2}}(W^+_{\mu}-W^-_{\mu}),\frac{1}
{\cos{\theta_W}}Z_{\mu}\right)
\end{array}
\end{eqnarray}
and the field-strenght tensors
\begin{eqnarray}\label{a3}
\begin{array}{c}
\displaystyle F_{\mu\nu} = \partial_{\mu}A_{\nu}
-   \partial_{\nu}A_{\mu}, \\ \\
\displaystyle W_{\mu\nu}^i = \partial_{\mu}W^i_{\nu}
-   \partial_{\nu}W^i_{\mu}, \\ \\
\displaystyle \tilde{F}_{\mu\nu} = \frac{1}{2}
\epsilon_{\mu\nu\rho\sigma}F^{\rho\sigma}.
\end{array}
\end{eqnarray}
The scale $\Lambda$ is introduced to keep the coupling constant
$a_i$ dimensionless \cite{c3}. In practice,
the $\Lambda$ are specified in the frame of the chosen model for
"new physics" that supports anomalous quartic
gauge boson couplings. In our case $\Lambda$ are fixed by value
of $M_W$ ($\sim 80$ GeV).
As one can see the operators ${\cal L}_0$ and ${\cal L}_c$
are $C$-, $P$-, $CP$-invariant. ${\cal L}_n$
violates both $C$- and  $CP$-invariance.
$\tilde{{\cal L}}_0$ is the $P$- and $CP$-violating operator.
$\tilde{{\cal L}}_n$ conserves $CP$-invariance
but violates $C$- and $P$-invariance separately.

Now we can obtain Feynman rules for anomalous quartic gauge
boson couplings.
\begin{center}
\begin{picture}(100,120)(0,0)\label{p1}
\Photon(0,0)(50,50){2}{4}\put(10,0){$A_{\mu},k_1$}
\Vertex(50,50){1.2}
\Photon(100,0)(50,50){2}{4} \put(110,0){$W_{\alpha}^+,p_3$}
\Photon(0,100)(50,50){2}{4}\put(10,100){$A_{\nu},k_2$}
\Photon(100,100)(50,50){2}{4}  \put(110,100){$W_{\beta}^-,p_4$}
\end{picture}
\end{center}
\begin{eqnarray}\label{a4}
\begin{array}{c}
\displaystyle = \frac{e^2}{8\Lambda^2}\left(4a_0g^{\alpha\beta}
((k_1k_2)g^{\mu\nu}-k_1^{\nu}k_2^{\mu})+a_c((k_1^{\alpha}k_2^{\beta}
+k_1^{\beta}k_2^{\alpha})g^{\mu\nu}
+(k_1k_2)(g^{\mu\alpha}
g^{\nu\beta}+g^{\nu\alpha}g^{\mu\beta})- \right. \\     \\
\displaystyle\left. - k_1^{\nu}(k_2^{\beta}g^{\mu\alpha}+k_2^{\alpha}
g^{\mu\beta})- k_2^{\nu}(k_1^{\beta}g^{\mu\alpha}+k_1^{\alpha}
g^{\mu\beta}))+4\tilde{a}_0g^{\alpha\beta}k_{1\rho}k_{2\sigma}\
\epsilon^{\mu\rho\nu\sigma}\right).
\end{array}
\end{eqnarray}
\begin{center}
\begin{picture}(100,100)(0,0)\label{p2}
\Photon(0,0)(50,50){2}{4}\put(10,0){$A_{\mu},k_1$}
\Vertex(50,50){1.2}
\Photon(100,0)(50,50){2}{4} \put(110,0){$W_{\alpha}^+,p_3$}
\Photon(0,100)(50,50){2}{4}\put(10,100){$Z_{\nu},k_2$}
\Photon(100,100)(50,50){2}{4}  \put(110,100){$W_{\beta}^-,p_4$}
\end{picture}
\end{center}
\begin{eqnarray}\label{a5}
\begin{array}{c}
\displaystyle = \frac{e^2}{16\Lambda^2\cos{\theta_W}}
\left(a_n(-(k_1k_2)(g^{\mu\alpha}g^{\nu\beta}-g^{\mu\beta}
g^{\nu\alpha})-(k_1p_3)(g^{\mu\beta}g^{\nu\alpha}
-g^{\mu\nu}g^{\alpha\beta}) - \right.\\ \\
-(k_1p_4)(g^{\mu\nu}g^{\alpha\beta}-g^{\mu\alpha}g^{\nu\beta})+ \\  \\
\displaystyle + k_2^{\mu}(k_1^{\alpha}g^{\nu\beta}-k_1^{\beta}
g^{\nu\alpha})+ p_3^{\mu}(k_1^{\beta}g^{\nu\alpha}-k_1^{\beta}
g^{\nu\alpha})+p_4^{\mu}(k_1^{\nu}g^{\beta\alpha}-k_1^{\alpha}
g^{\nu\beta})- \\ \\
\displaystyle -g^{\mu\nu}(k_1^{\beta}p_3^{\alpha}-k_1^{\alpha}
p_4^{\beta}) - g^{\mu\alpha}(k_1^{\nu}p_4^{\beta}-k_1^{\beta}
k_2^{\nu}) -g^{\mu\beta}(k_1^{\alpha}k_2^{\nu}-k_1^{\nu}
k_3^{\alpha}))+ \\ \\
\tilde{a}_nk_{1\rho}((k_1+k_2)^{\nu}\epsilon^{\alpha\beta\mu\rho}
+(k_1+p_3)^{\alpha}\epsilon^{\beta\nu\mu\rho} + (k_1+p_4)^{\beta}
\epsilon^{\nu\alpha\mu\rho} - \\ \\
-(k_2-k_1)_{\sigma}g^{\nu\alpha}\epsilon^{\sigma\beta\mu\rho}
-(p_3-p_4)_{\sigma}g^{\beta\alpha}\epsilon^{\sigma\nu\mu\rho}
-(p_4-k_2)_{\sigma}g^{\nu\beta}\epsilon^{\sigma\alpha\mu\rho}).
\end{array}
\end{eqnarray}

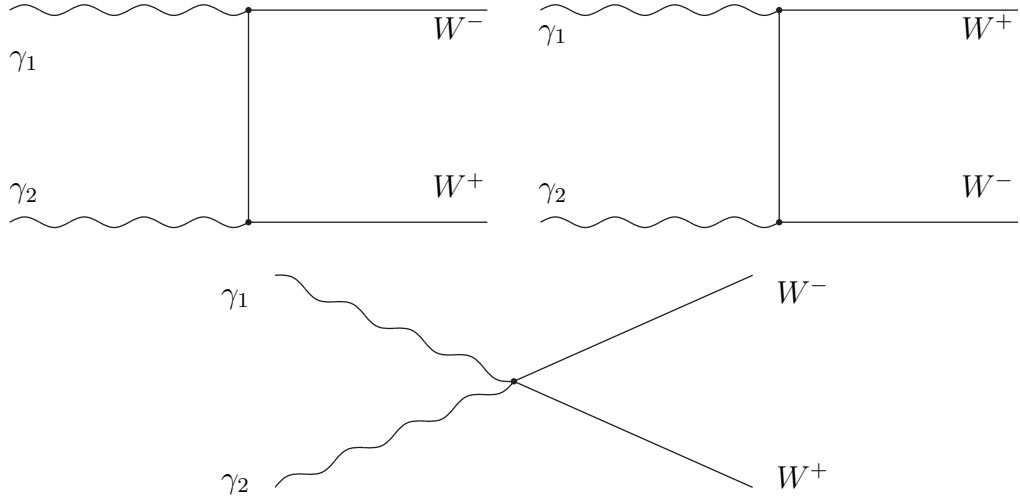
\begin{figure}[h]
\begin{center}
\begin{picture}(400,200)(0,0)

%1
\Photon(10,110)(100,110){2}{4}\put(10,120){$\gamma_2$}
\Vertex(100,110){1.2}
\Line(100,110)(190,110)\put(170,120){$W^+$}
\Line(100,110)(100,190)
\Photon(10,190)(100,190){2}{4}\put(10,170){$\gamma_1$}
\Vertex(100,190){1.2}
\Line(100,190)(190,190)\put(170,180){$W^-$}

%2
\Photon(210,110)(300,110){2}{4}\put(210,120){$\gamma_2$}
\Vertex(300,110){1.2}
\Line(300,110)(390,110)\put(370,120){$W^-$}
\Line(300,110)(300,190)
\Photon(210,190)(300,190){2}{4}\put(210,180){$\gamma_1$}
\Vertex(300,190){1.2}
\Line(300,190)(390,190)\put(370,180){$W^+$}

%3
\Photon(110,10)(200,50){2}{4}\put(90,10){$\gamma_2$}
\Vertex(200,50){1.2}
\Photon(110,90)(200,50){2}{4}\put(90,80){$\gamma_1$}
\Line(200,50)(290,10)\put(300,10){$W^+$}
\Line(200,50)(290,90)\put(300,80){$W^-$}

\end{picture}
\end{center}
\caption{The Feynman diagrams for $W^+W^-$-production}\label{p3}
\end{figure}

\begin{figure}[h!]
\begin{center}
\begin{picture}(400,400)(0,0)

%1
\Photon(10,370)(100,370){2}{4}\put(10,360){$\gamma_1$}
\Photon(10,310)(100,310){2}{4}\put(10,320){$\gamma_2$}
\Vertex(100,370){1.2}
\Line(100,370)(145,370)
\Vertex(145,370){1.2}
\Line(145,370)(190,390)\put(170,370){$W^+$}
\Line(145,370)(190,350)\put(170,340){$Z$}
\Line(100,310)(100,370)
\Vertex(100,310){1.2}
\Line(100,310)(190,310)\put(170,315){$W^-$}
\put(100,295){$1$}

%2
\Photon(210,370)(300,370){2}{4}\put(210,360){$\gamma_1$}
\Photon(210,310)(300,310){2}{4}\put(210,320){$\gamma_2$}
\Vertex(300,370){1.2}
\Line(300,370)(345,370)
\Vertex(345,370){1.2}
\Line(345,370)(390,390)\put(370,370){$W^-$}
\Line(345,370)(390,350)\put(370,340){$Z$}
\Line(300,310)(300,370)
\Vertex(300,310){1.2}
\Line(300,310)(390,310)\put(370,315){$W^+$}
\put(300,295){$2$}

%3
\Photon(110,280)(180,280){2}{4}\put(110,270){$\gamma_1$}
\Photon(110,220)(180,220){2}{4}\put(110,230){$\gamma_2$}
\Vertex(180,280){1.2}
\Line(180,280)(280,300)\put(270,280){$W^+$}
\Line(180,280)(220,250)
\Vertex(220,250){1.2}
\Line(220,250)(290,250)\put(295,245){$Z$}
\Line(180,220)(220,250)
\Vertex(180,220){1.2}
\Line(180,220)(280,200)\put(270,210){$W^-$}
\put(200,200){$3$}

%4
\Photon(10,190)(100,190){2}{4}\put(10,180){$\gamma_1$}
\Photon(10,130)(100,130){2}{4}\put(10,140){$\gamma_2$}
\Vertex(100,190){1.2}
\Line(100,190)(190,190)\put(170,180){$W^-$}
\Line(100,130)(100,190)
\Vertex(100,130){1.2}
\Line(100,130)(190,150)\put(170,150){$W^+$}
\Line(100,130)(190,110)\put(170,120){$Z$}
\put(100,100){$4$}

%5
\Photon(210,190)(300,190){2}{4}\put(210,180){$\gamma_1$}
\Photon(210,130)(300,130){2}{4}\put(210,140){$\gamma_2$}
\Vertex(300,190){1.2}
\Line(300,190)(390,190)\put(370,180){$W^+$}
\Line(300,130)(300,190)
\Vertex(300,130){1.2}
\Line(300,130)(390,150)\put(370,150){$W^-$}
\Line(300,130)(390,110)\put(370,120){$Z$}
\put(300,100){$5$}

%6
\Photon(10,90)(100,50){2}{4}\put(30,90){$\gamma_1$}
\Photon(10,10)(100,50){2}{4}\put(30,10){$\gamma_2$}
\Vertex(100,50){1.2}
\Line(100,50)(190,10)\put(150,10){$W^-$}
\Line(100,50)(145,65)
\Vertex(145,65){1.2}
\Line(145,65)(190,90)\put(150,80){$W^+$}
\Line(145,65)(190,60)\put(150,50){$Z$}
\put(100,0){$6$}

%7
\Photon(210,90)(300,50){2}{4}\put(230,90){$\gamma_1$}
\Photon(210,10)(300,50){2}{4}\put(230,10){$\gamma_2$}
\Vertex(300,50){1.2}
\Line(300,50)(390,10)\put(350,10){$W^+$}
\Line(300,50)(345,65)
\Vertex(345,65){1.2}
\Line(345,65)(390,90)\put(350,80){$W^-$}
\Line(345,65)(390,60)\put(350,50){$Z$}
\put(300,0){$7$}

\end{picture}
\end{center}
\caption{The first-order Feynman diagrams
           for $W^+W^-Z$-production}\label{p4}
\end{figure}
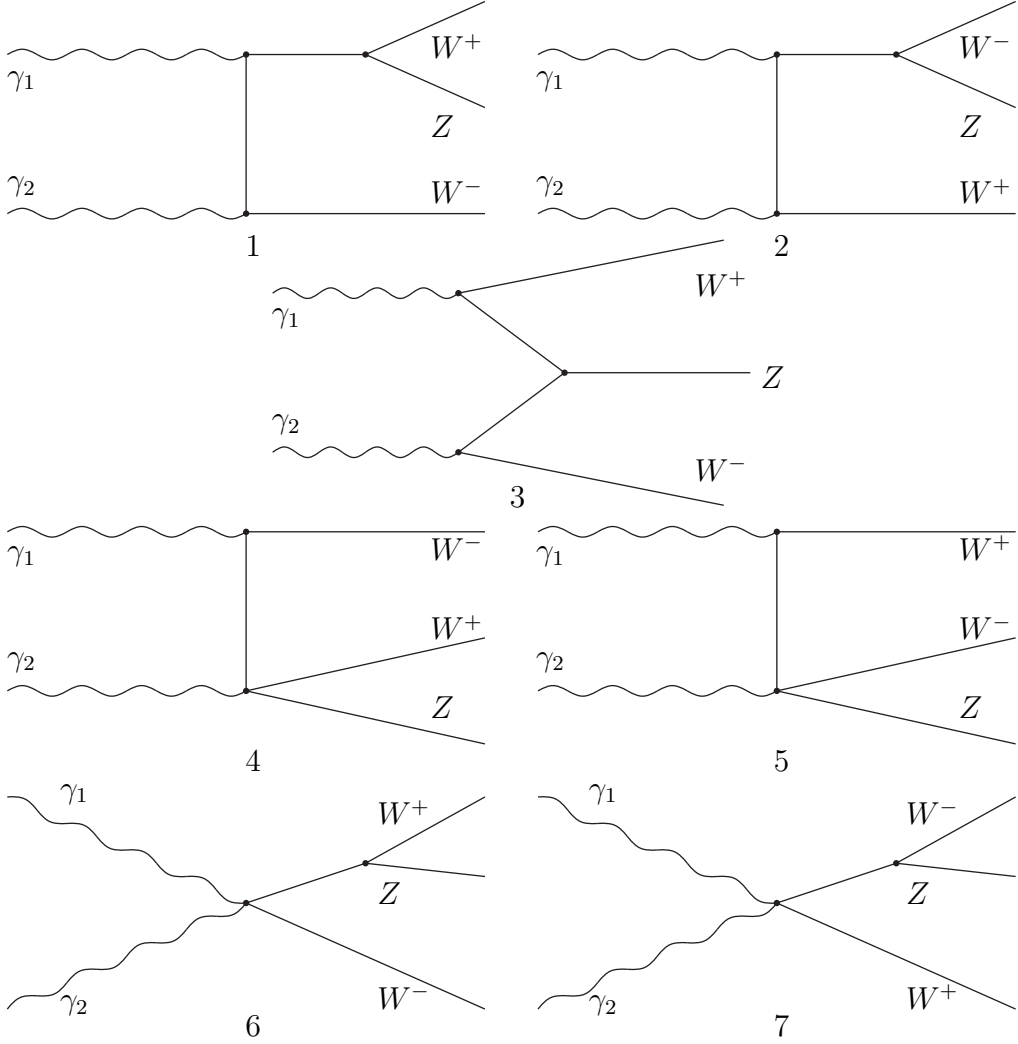

All momenta of bosons are supposed as incoming.
Feynman diagrams for the $\gamma\gamma\rightarrow W^+W^-$ and
$\gamma\gamma\rightarrow W^+W^-Z$ processes involving quartic
gauge boson couplings
are presented on Fig.\ref{p3} and Fig.\ref{p4}.

%%%%%%%%%%%%%%%%%%%%%%%%%%%%%%%%%%%%%%%%%%%%%%%%%%%%%%%%%%%%%%%%%%
%%% 1.1
%%%%%%%%%%%%%%%%%%%%%%%%%%%%%%%%%%%%%%%%%%%%%%%%%%%%%%%%%%%%%%%%%%

\newpage
%%%%%%%%%%%%%%%%%%%%%%%%%%%%%%%%%%%%%%%%%%%%%%%%%%%%%%%%%%%%%%%%%%
For the SM triple and quartic gauge boson couplings see for example
\cite{c9}.

\section{Numerical Results}

$ $

In this section we consider dependence of the cross sections
for $\gamma\gamma\rightarrow W^+W^-$ and
$\gamma\gamma\rightarrow W^+W^-Z$ on the five anomalous couplings studied in the Section $2$.

We start from explicit expression for the
amplitude of the process $\gamma\gamma\rightarrow W^+W^-$ 
\begin{align}
M = & G_v \epsilon_{\mu}(k_1)\epsilon_{\nu}(k_2)
\epsilon_{\alpha}(p_{+})\epsilon_{\beta}(p_{-})
%\epsilon_{\gamma}(k_3) 
M_T^{\mu\nu\alpha\beta},
\end{align}
where
\begin{align}
M_T^{\mu\nu\alpha\beta} = \sum_{i=1}^{3}
M_i^{\mu\nu\alpha\beta},
\end{align}
$k_1$, $k_2$, $p_{+}$, $p_{-}$ are four-momenta for the $\gamma$, $\gamma$, $W^{+}$, $W^{-}$ 
and $\epsilon_{\mu}(k_1), \epsilon_{\nu}(k_2),\epsilon_{\alpha}(p_{+}), \epsilon_{\beta}(p_{-})$ 
-- polarizations, respectively.
$$G_v = e^3 \cot\theta_W.$$
It is convenient to define the triple-gauge-boson coupling as
\begin{align}\label{a1}
{\Gamma}_{3}^{\alpha\beta\gamma}(P_1,P_2) = & [(2P_1 +
P_2)^{\beta} g^{\alpha\gamma} - (2P_2 + P_1)^{\alpha}
g^{\beta\gamma} + (P_2 - P_1)^{\gamma} g^{\alpha\beta}],
\end{align}
the SM quartic-gauge-boson coupling as
\begin{align}\label{a2}
{\Gamma}_{4}^{\mu\nu\alpha\beta} = & g^{\mu\alpha} g^{\nu\beta} +
g^{\mu\beta} g^{\nu\alpha} -2 g^{\mu\nu} g^{\alpha\beta},
\end{align}
and the propagator tensor for the vector boson     as
\begin{align}\label{a3}
D^{\mu\nu}(k) = & \frac{(g{\mu\nu} - k^{\mu}k^{\nu}/m^2)}{k^2 -
m^2}.
\end{align}
Anomalous quartic  boson couplings are defined in (\ref{a4}) and (\ref{a5}).
Using the expression defined above, the contributions of the three
Feynman diagrams (see fig. \ref{p3}) for the $WW$-production can be written as
\begin{align}\label{b8}
M_{1}^{\mu\nu\alpha\beta} = & {\Gamma}_{3}^{\mu\alpha\xi}
(-k_1,p_{+}) D_{\xi\lambda}(p_+ -
k_1){\Gamma}_{3}^{\nu\beta\lambda}(-k_2,p_-),
\end{align}
\begin{align}\label{b9}
M_{2}^{\mu\nu\alpha\beta} = & {\Gamma}_{3}^{\mu\beta\xi}
(-k_1,p_{-}) D_{\xi\lambda}(p_- -
k_1){\Gamma}_{3}^{\nu\alpha\lambda}(-k_2,p_+),
\end{align}
\begin{align}\label{b10}
M_{3}^{\mu\nu\alpha\beta} = & {\Gamma}_{4}^{\mu\nu\alpha\beta}.
\end{align}

\begin{figure}[h!]
\begin{minipage}[b]{.975\linewidth}
\centering
\includegraphics[width=\linewidth, height=3.8in, angle=0]{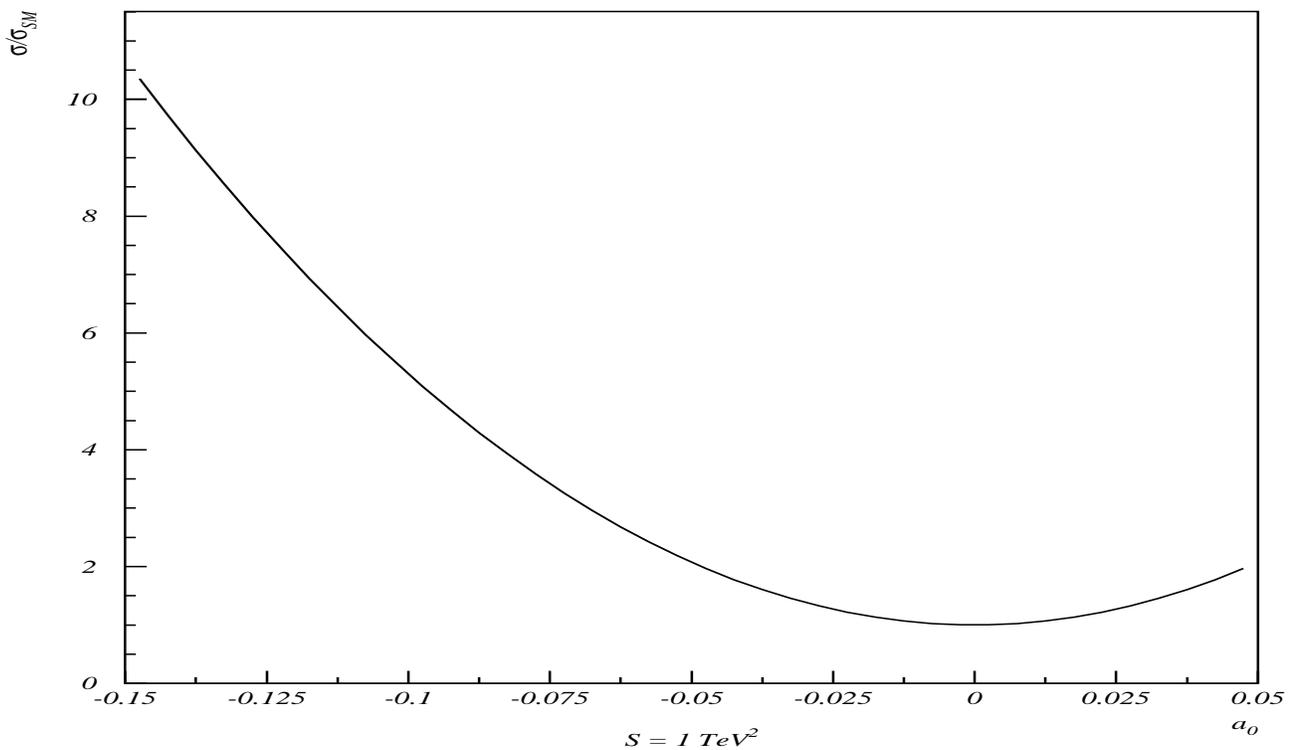}
\caption{ The effect of the anomalous $a_0$
coupling on the cross section $\sigma(W^+W^-Z)$,
normalised to the SM values } \label{p9}
\end{minipage}\hfill
\end{figure}

\begin{figure}[h!]
\begin{minipage}[b]{.975\linewidth} \centering
\includegraphics[width=\linewidth, height=3.8in, angle=0]{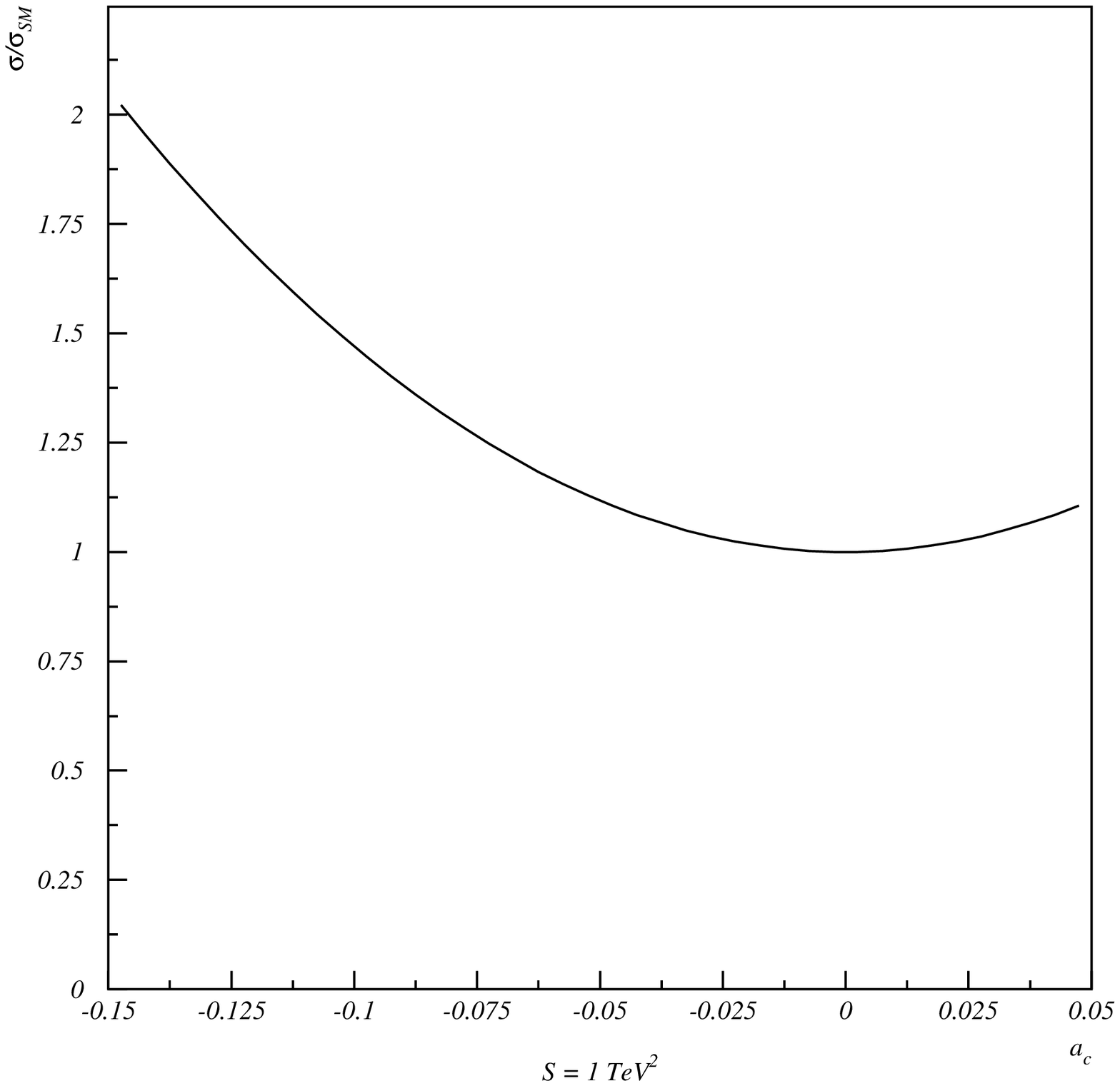}
\caption{The effect of the anomalous $a_c$ coupling
on the cross section $\sigma(W^+W^-Z)$, normalised
to the SM values } \label{p10}
\end{minipage}
\end{figure}

\begin{figure}[h!]
\begin{minipage}[b]{.975\linewidth}
\centering
\includegraphics[width=\linewidth, height=3.8in, angle=0]{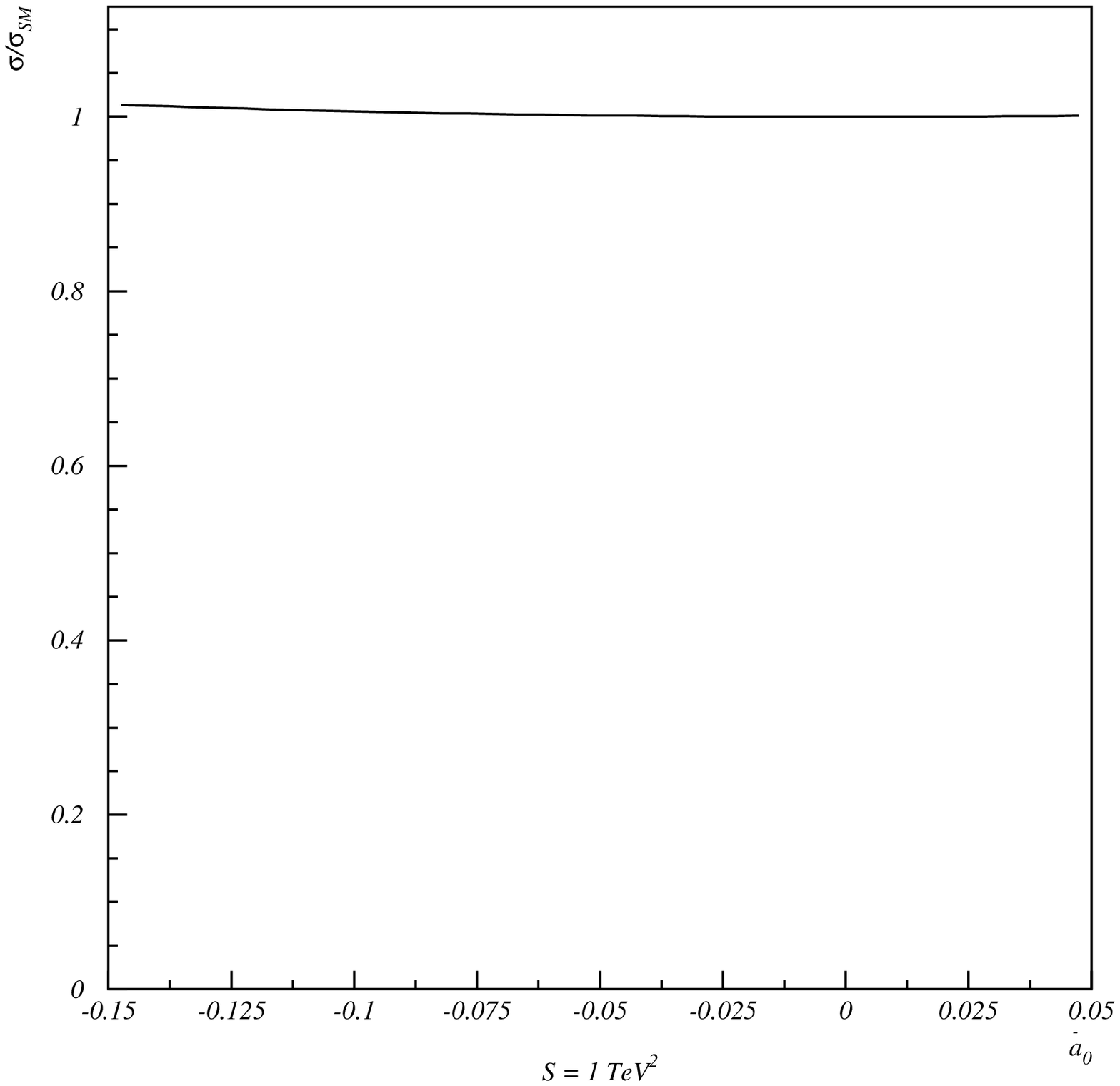}
\caption{The effect of the anomalous $\tilde{a}_0$
coupling on the cross section $\sigma(W^+W^-Z)$,
normalised to the SM values} \label{p11}
\end{minipage}\hfill
\end{figure}

\begin{figure}[h!]
\begin{minipage}[b]{.975\linewidth}
\centering
\includegraphics[width=\linewidth, height=3.8in, angle=0]{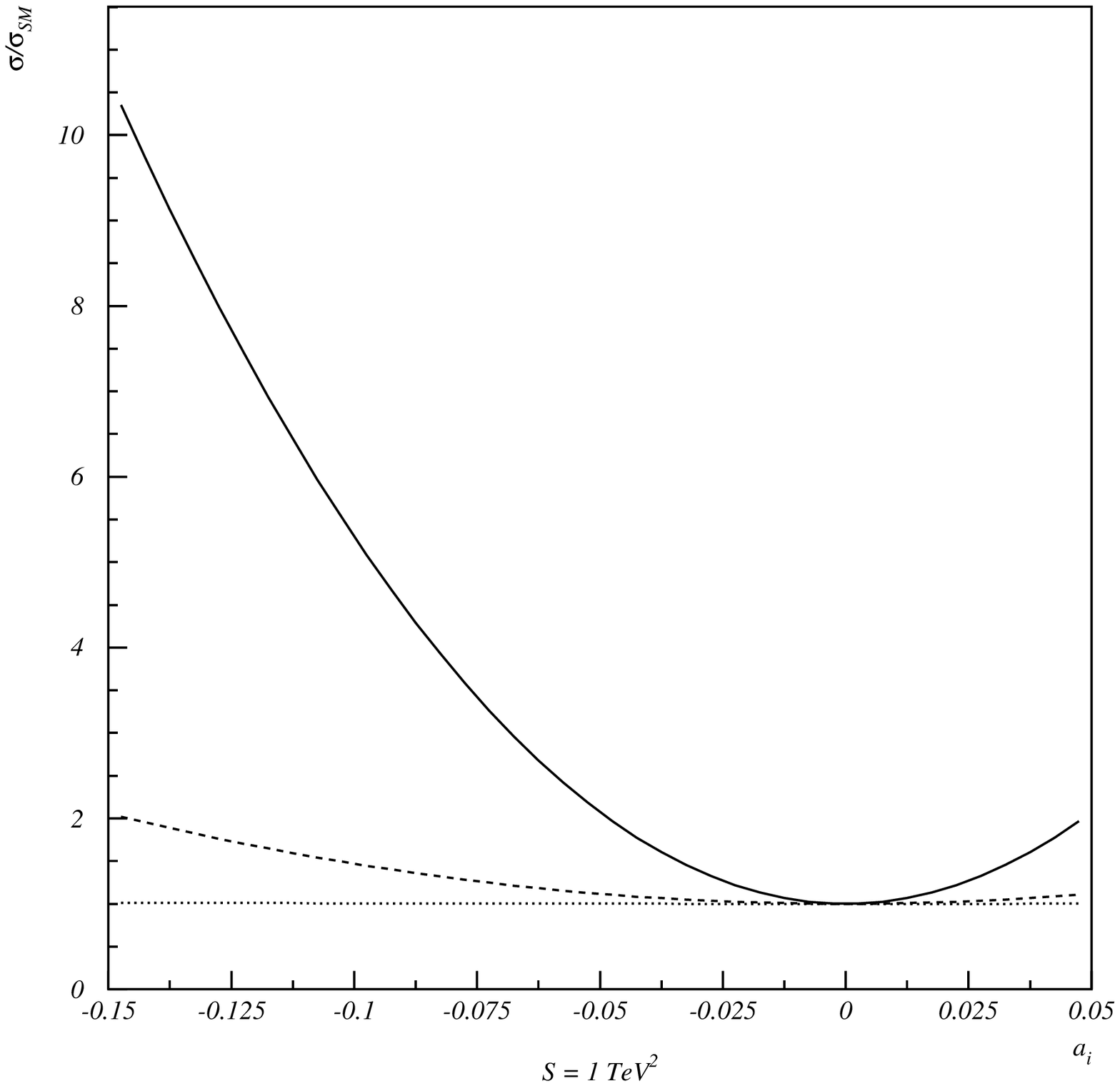}
\caption{Comparison of the dependences of the cross sections
$\sigma(W^+W^-Z)$. Solid line presents $a_0$-, scratched line -
$a_c$-, dotted line  $\tilde{a}_0$-dependence} \label{p12}
\end{minipage}
\end{figure}

The expressions of aplitude for a triple boson production (see fig. \ref{p4}) can be defined in the following way:
\begin{align}\label{b1}
M_{1}^{\mu\nu\alpha\beta\gamma} = & {\Gamma}_{3}^{\alpha\gamma\xi}
(p_{+},k_3) D_{\xi\sigma}(p_{+}+k_3) {\Gamma}_{3}^{\mu\sigma\rho}
(k_1,-(p_{+} + k_3)) \nonumber\\ & D_{\rho\lambda}(p_{-} - k_2)
{\Gamma}_{3}^{\beta\nu\lambda}(-p_{-},k_2) + [k_1\leftrightarrow
k_2; \mu\leftrightarrow\nu]
\end{align}
\begin{align}\label{b2}
M_{2}^{\mu\nu\alpha\beta\gamma} = & {\Gamma}_{3}^{\gamma\beta\xi}
(k_3,p_{-}) D_{\xi\sigma}(p_{-}+k_3) {\Gamma}_{3}^{\sigma\nu\rho}
(-(p_{-} + k_3),k_2) \nonumber\\ & D_{\rho\lambda}(k_1 - p_{+})
{\Gamma}_{3}^{\alpha\mu\lambda}(-p_{+},k_1) + [k_1\leftrightarrow
k_2; \mu\leftrightarrow\nu]
\end{align}
\begin{align}\label{b3}
M_{3}^{\mu\nu\alpha\beta\gamma} = & {\Gamma}_{3}^{\mu\alpha\xi}
(k_1,-p_{+}) D_{\xi\sigma}(k_1 - p_{+})
{\Gamma}_{3}^{\gamma\sigma\rho} (-k_3,(k_1 - p_{+})) \nonumber\\ &
D_{\rho\lambda}(p_{-} - k_2)
{\Gamma}_{3}^{\nu\beta\lambda}(-k_2,p_{-}) + [k_1\leftrightarrow
k_2; \mu\leftrightarrow\nu]
\end{align}
\begin{align}\label{b4}
M_{4}^{\mu\nu\alpha\beta\gamma} = & {\Gamma}_{3}^{\beta\nu\xi}
(-p_{-},k_2) D_{\xi\lambda}(k_2 - p_{-})
{\Gamma}_{4}^{\lambda\alpha\mu\gamma} + [k_1\leftrightarrow k_2;
\mu\leftrightarrow\nu]
\end{align}
\begin{align}\label{b5}
M_{5}^{\mu\nu\alpha\beta\gamma} = & {\Gamma}_{3}^{\mu\alpha\xi}
(k_1,-p_{+}) D_{\xi\lambda}(k_1 - p_{+})
{\Gamma}_{4}^{\lambda\beta\nu\gamma} + [k_1\leftrightarrow k_2;
\mu\leftrightarrow\nu]
\end{align}
\begin{align}\label{b6}
M_{6}^{\mu\nu\alpha\beta\gamma} = & {\Gamma}_{3}^{\alpha\gamma\xi}
(p_{+},k_3) D_{\xi\lambda}(k_3 + p_{+})
{\Gamma}_{4}^{\lambda\beta\nu\mu}
\end{align}
\begin{align}\label{b7}
M_{7}^{\mu\nu\alpha\beta\gamma} = & {\Gamma}_{3}^{\gamma\beta\xi}
(k_3,p_{-}) D_{\xi\lambda}(-k_3 - p_{-})
{\Gamma}_{4}^{\lambda\alpha\nu\mu}
\end{align}
where $[k_1\leftrightarrow k_2; \mu\leftrightarrow\nu]$ defines
the crossed contributions of the initial photons.
The covariant formulae of amplitudes for considering processes are also
presented in \cite{c8,c9}.
\begin{figure}[h!]
\leavevmode
\begin{minipage}[b]{.975\linewidth} \centering
\centering
\includegraphics[width=\linewidth, height=3.8in, angle=0]{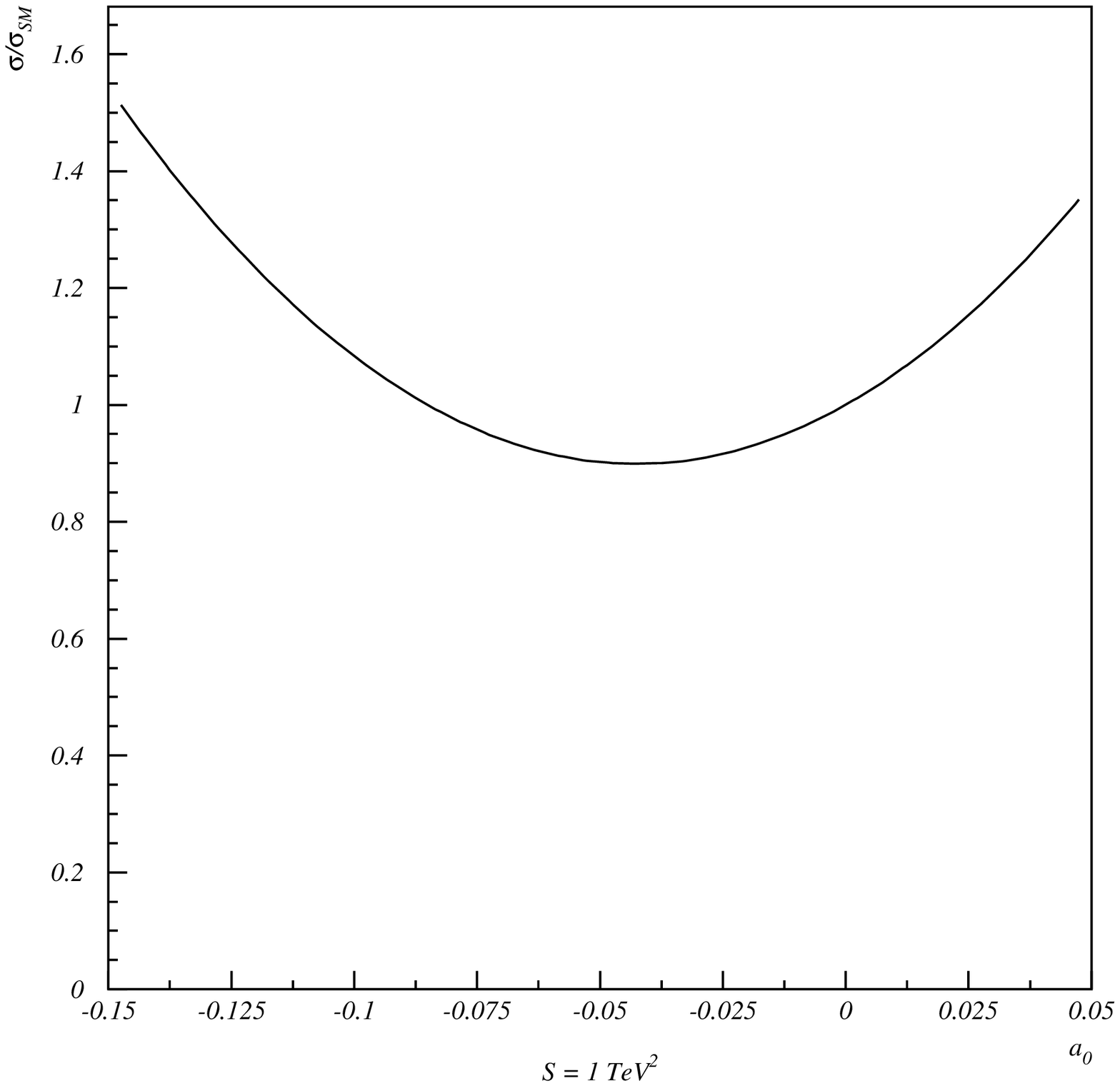}
\caption{ The effect of the anomalous $a_0$ coupling on the cross
section $\sigma(W^+W^-)$, normalised to the SM values } \label{p5}
\end{minipage}\hfill
\end{figure}
\begin{figure}[h!]
\begin{minipage}[b]{.975\linewidth} \centering
\includegraphics[width=\linewidth, height=3.8in, angle=0]{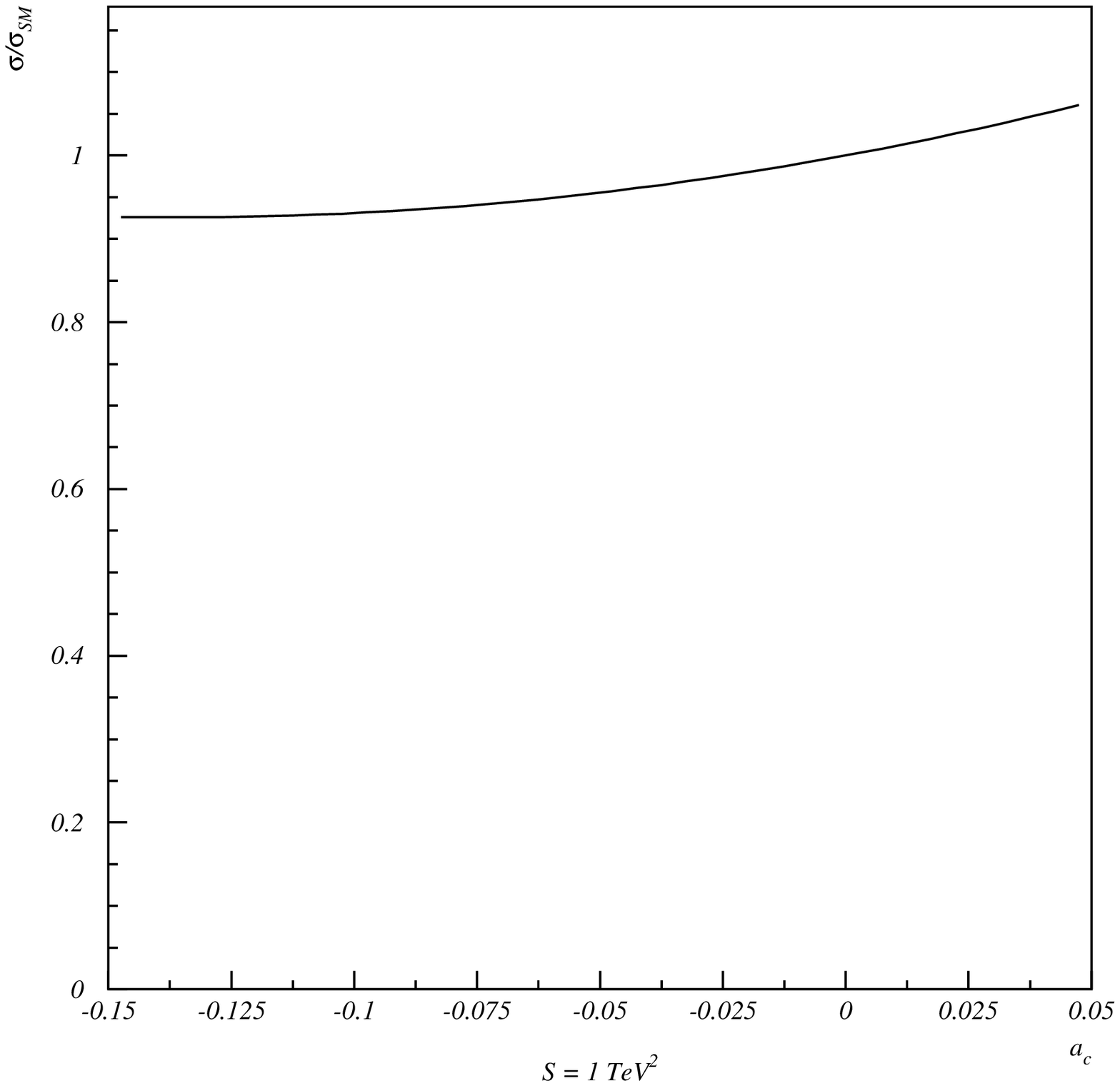}
\caption{The effect of the anomalous $a_c$ coupling on the cross
section $\sigma(W^+W^-)$, normalised to the SM values } \label{p6}
\end{minipage}\hfill
\end{figure}
\begin{figure}[h!]
\begin{minipage}[b]{.975\linewidth}
\centering
\includegraphics[width=\linewidth, height=3.8in, angle=0]{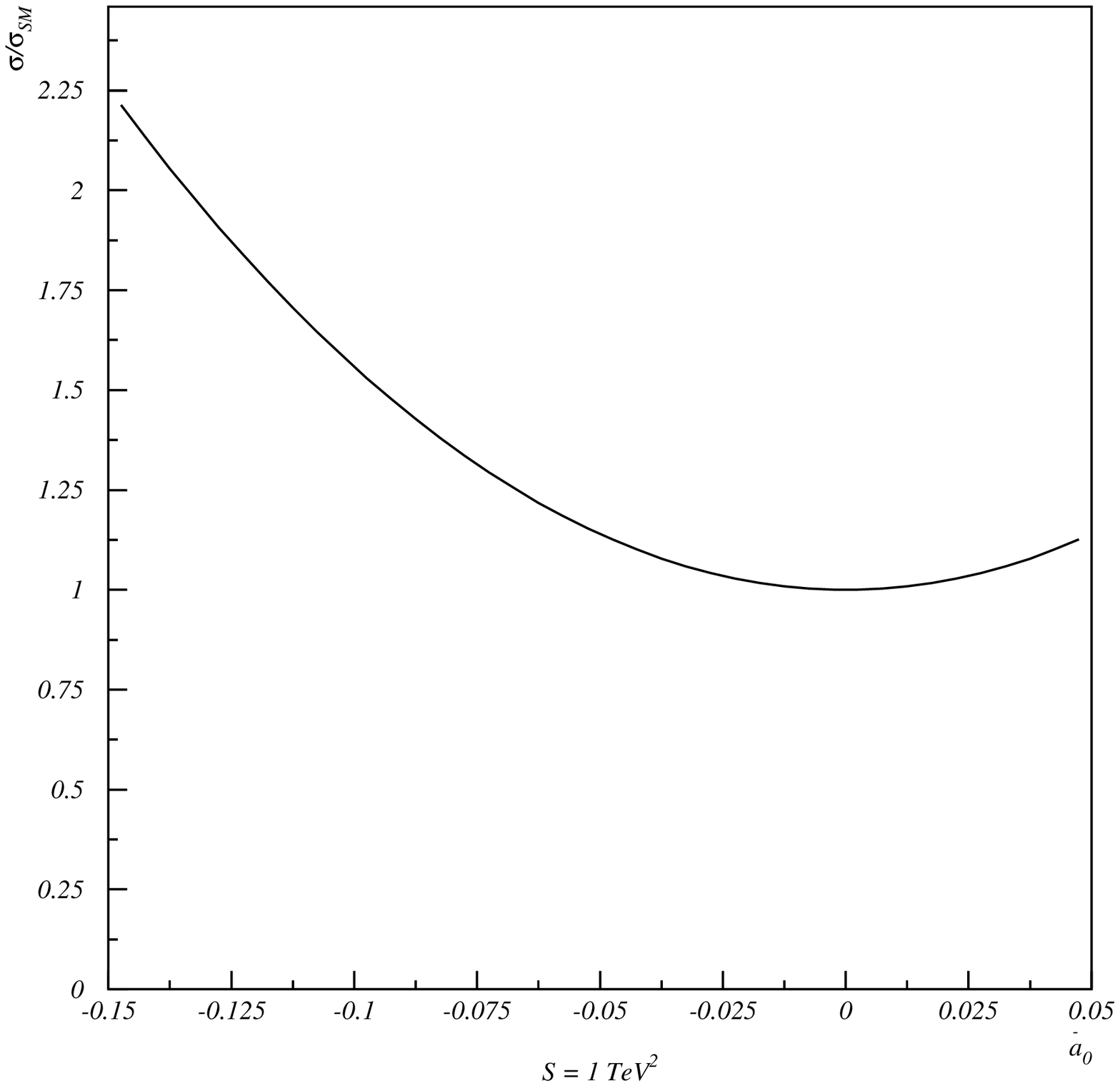}
\caption{The effect of the anomalous $\tilde{a}_0$ coupling on the
cross section $\sigma(W^+W^-)$, normalised to the SM values} \label{p7}
\end{minipage}\hfill
\end{figure}
\begin{figure}[h!]
\begin{minipage}[b]{.975\linewidth}
\centering
\includegraphics[width=\linewidth, height=3.8in, angle=0]{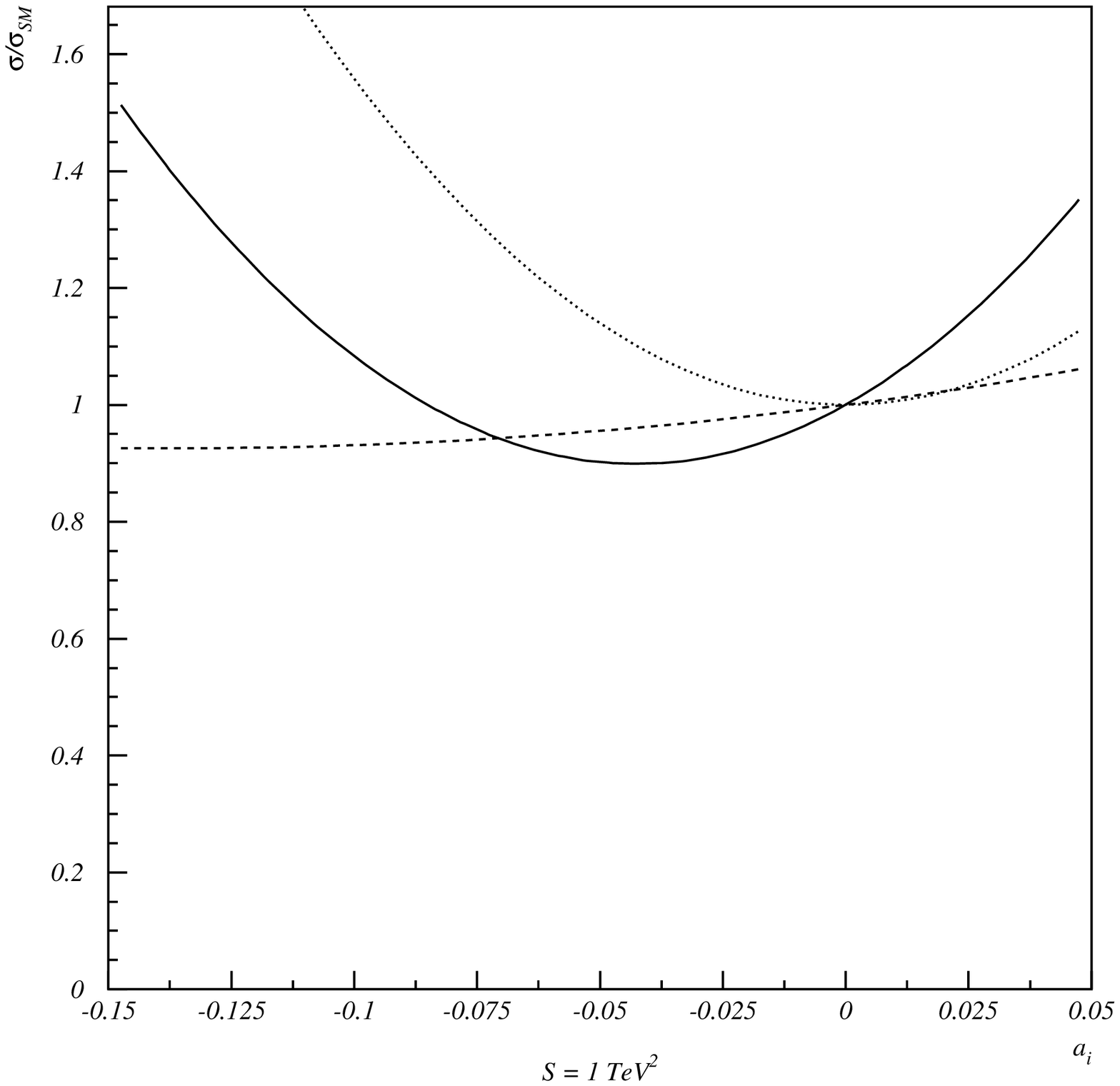}
\caption{The comparison of the dependences of the cross sections
$\sigma(W^+W^-)$. Solid line presents $a_0$-, scratched line
$a_c$-, dotted line  $\tilde{a}_0$-dependence} \label{p8}
\end{minipage} \hfill
\end{figure}
Using the Monte-Carlo method of integration \cite{c7}
we obtain the cross section of
$\gamma\gamma\rightarrow W^+W^-$
at various values of $\sqrt{S}$
\begin{eqnarray}\label{a6}
\begin{array}{c}
\displaystyle\sigma = \sum\limits_{\lambda_1\lambda_2\lambda_3
\lambda_4}\int|M_{\lambda_1\lambda_2\lambda_3\lambda_4}|^2
d\Gamma,
\end{array}
\end{eqnarray}
where $M_{\lambda_1\lambda_2\lambda_3\lambda_4}$ is the total
amplitude of the process $\gamma\gamma\rightarrow W^+W^-$.
$\lambda_1,\lambda_2,\lambda_3,\lambda_4$
are polarizations of initial photons and final bosons respectively.
The cross section $\sigma$ for  $W^+W^-$ production at
$\sqrt{S}=1$TeV is $\sim 110${\it pb} and for
$W^+W^-Z$ production is $\sim 12${\it pb}. We consider the following
experimental conditions: unpolarized $\gamma\gamma$ beams at
$\sqrt{S}=1$ TeV with luminosity ${\cal L}=100fb^{-1}/year$
according to
TESLA TDR \cite{c10}. Note that $\sigma(W^+W^-)$ depends on
$a_0$,  $a_c$ and $\tilde{a}_0$, but $\sigma(W^+W^-Z)$ has additional
dependence on $a_n$ and $\tilde{a}_n$. Figures \ref{p5} --
\ref{p8} shows dependence of three total cross sections
$\sigma(W^+W^-)$ on anomalous parameters.
The cross section depends on anomalous couplings quadratically.
The fact that the minima of the curves are close to the SM point
$a_i=0$ means that the interference between the anamolous and
the standard part of the
matrix element is very small. In the region for $a_i$ about
$0.5$ the anomalous part of the cross section achieves the
values about $2\sigma_{SM}\sim 3\sigma_{SM}$.
That stands for the evidence of great sensitivity of the
cross section $\sigma(W^+W^-)$ to anomalous couplings.
In figures \ref{p9} -- \ref{p12} the dependences of
the totall cross sections $\sigma(W^+W^-Z)$
for $W^+W^-Z$ production are shown.
Comparing two sets of figures on can conclude that the $W^+W^-Z$
production is more suitable (more sensitive) for analysing
the anomalous $a_0$ coupling in case of
unpolarized $\gamma\gamma$
collisions
while $\gamma\gamma\rightarrow W^+W^-$ has greater abilites
then previous  process for investigation of $a_c$ and
$\tilde{a}_0$. The last anomalous coupling $\tilde{a}_0$ aren't
examined by the investigation of
$\gamma\gamma\rightarrow W^+W^-Z$ at all. The sensibilites
of unpolarized $e^+e^-\rightarrow W^+W^-$ and
$e^+e^-\rightarrow W^+W^-Z$ for considered anomalous
couplings \cite{c5},\cite{c3} is less in comparison with $\gamma\gamma$
beams by about ten times.

\begin{figure}[h!]
\begin{minipage}[b]{.975\linewidth}
\centering
\includegraphics[width=\linewidth, height=3.8in, angle=0]{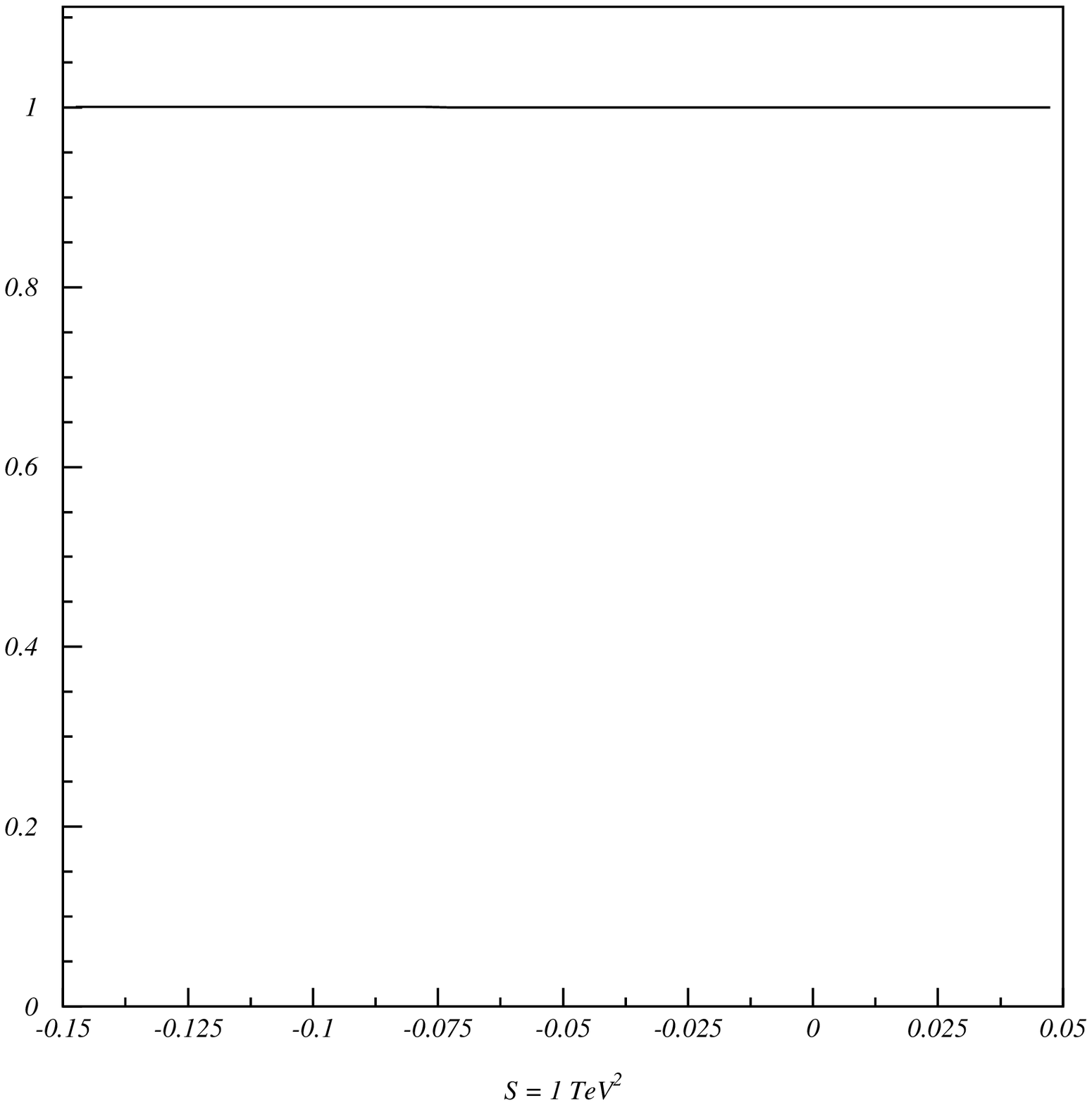}
\caption{ The effect of the anomalous $a_n$
coupling on the cross section $\sigma(W^+W^-Z)$,
normalised to the SM values } \label{p13}
\end{minipage}\hfill
\end{figure}

\begin{figure}[h!]
\begin{minipage}[b]{.975\linewidth} \centering
\includegraphics[width=\linewidth, height=3.8in, angle=0]{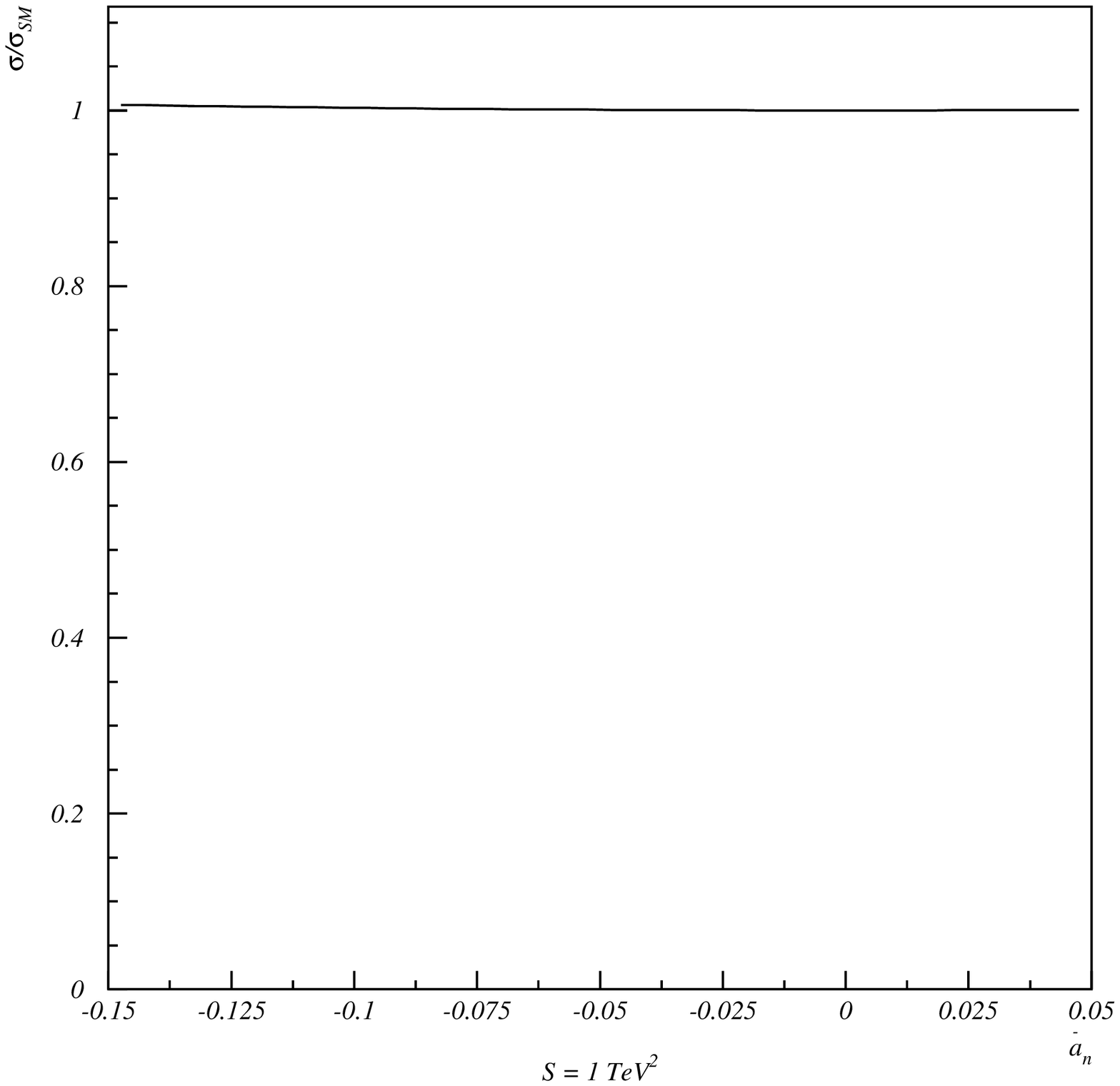}
\caption{The effect of the anomalous $\tilde{a}_n$
coupling on the cross section $\sigma(W^+W^-Z)$,
normalised to the SM values } \label{p14}
\end{minipage}\hfill
\end{figure}

The influences $a_n$ and $\tilde{a}_n$ to $\sigma(W^+W^-Z)$
are shown in figures \ref{p13} -- \ref{p14}.  They are negligible
in comparison with
$a_0$, $a_c$. The couplings $a_n$ and $\tilde{a}_0$
almost have the same order.


\begin{figure}[h!]
\begin{minipage}[b]{.975\linewidth}
\centering
\includegraphics[width=\linewidth, height=3.8in, angle=0]{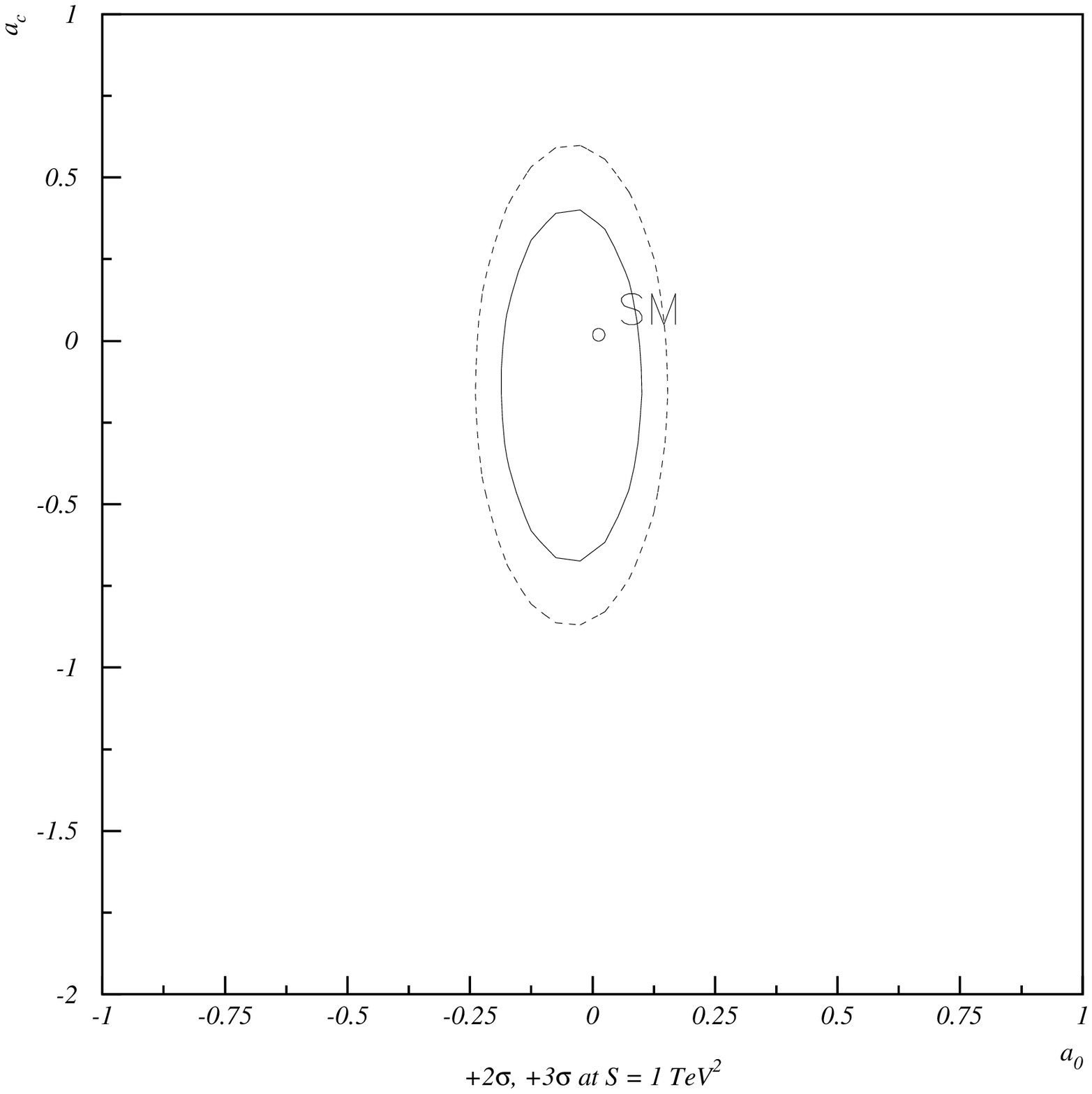}
\caption{ Contour plots for $+2\sigma$, $+3\sigma$ deviations
of $\sigma(W^+W^-)$ at $\int {\cal L} = 100fb^{-1}, \sqrt{S}=1TeV$ }
\label{p15}
\end{minipage}\hfill
\end{figure}

\begin{figure}[h!]
\begin{minipage}[b]{.975\linewidth} \centering
\includegraphics[width=\linewidth, height=3.8in, angle=0]{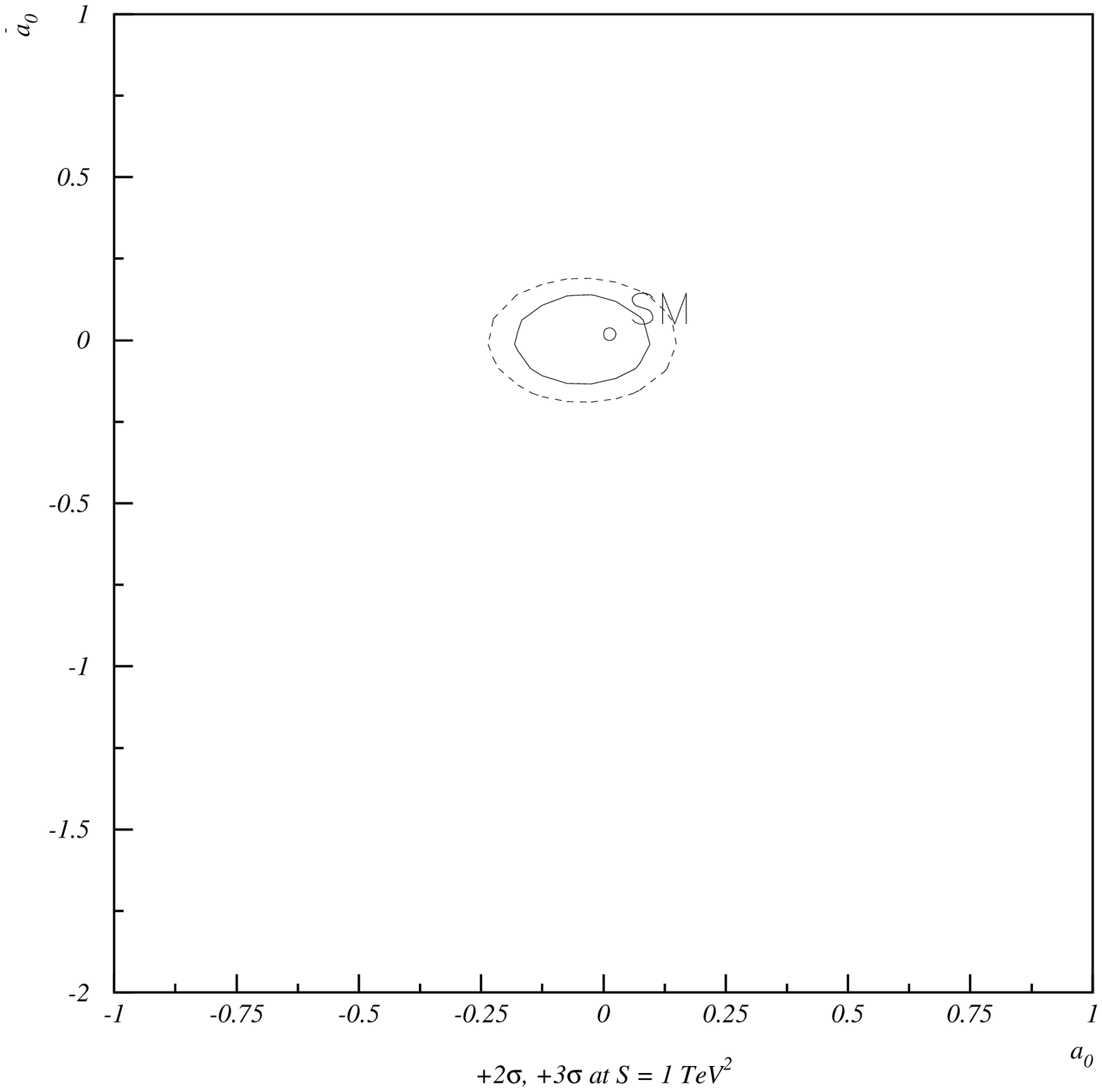}
\caption{Contour plots for $+2\sigma$, $+3\sigma$ deviations of
$\sigma(W^+W^-)$ at $\int {\cal L} = 100fb^{-1}, \sqrt{S}=1TeV$ }
\label{p16}
\end{minipage}\hfill
\end{figure}

\begin{figure}[h!]
\begin{minipage}[b]{.975\linewidth}
\centering
\includegraphics[width=\linewidth, height=3.8in, angle=0]{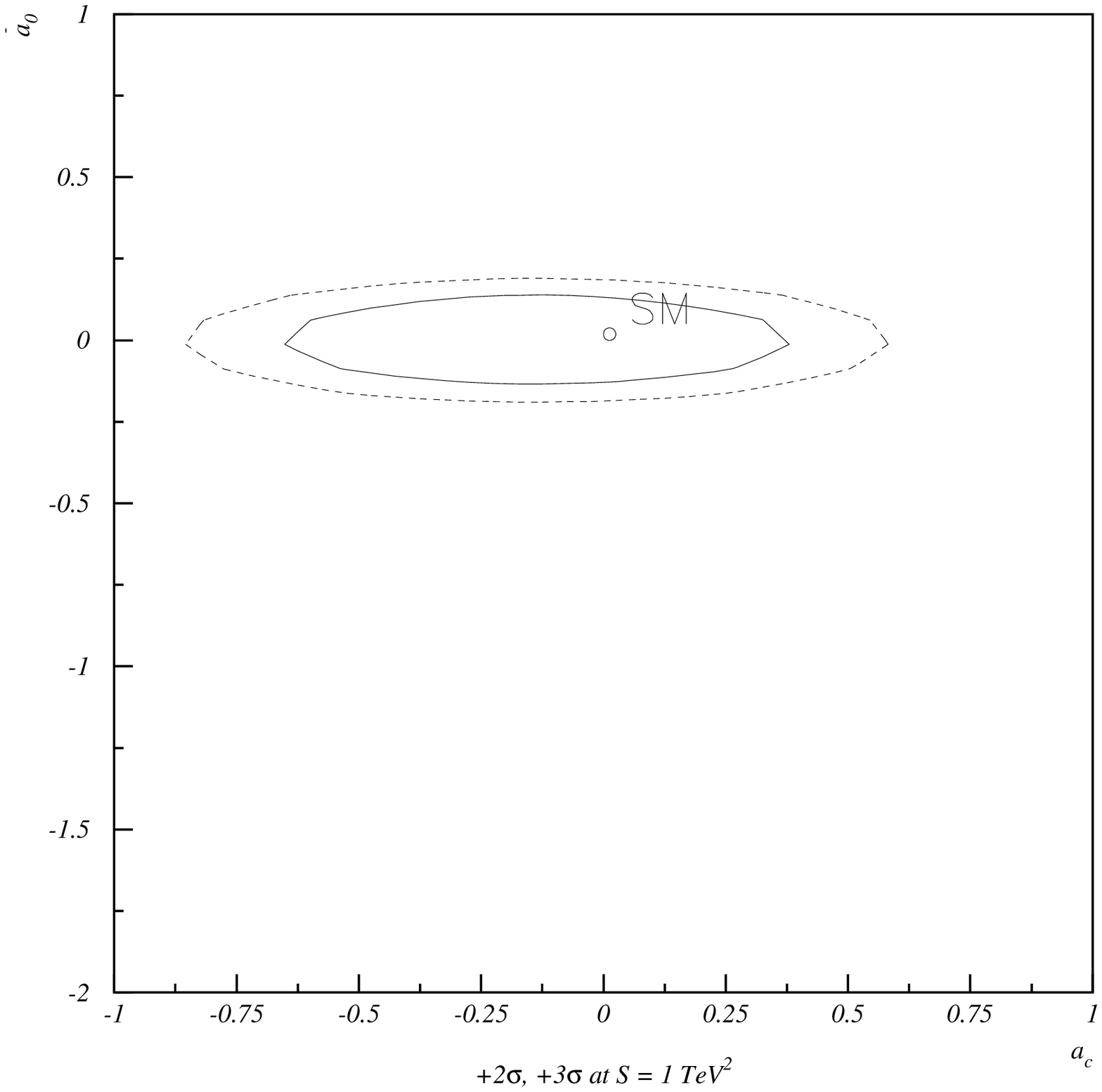}
\caption{ Contour plots for $+2\sigma$, $+3\sigma$ deviations of
$\sigma(W^+W^-)$ at $\int {\cal L} = 100fb^{-1}, \sqrt{S}=1TeV$ }
\label{p17}
\end{minipage}\hfill
\end{figure}

\begin{figure}[h!]
\begin{minipage}[b]{.975\linewidth} \centering
\includegraphics[width=\linewidth, height=3.8in, angle=0]{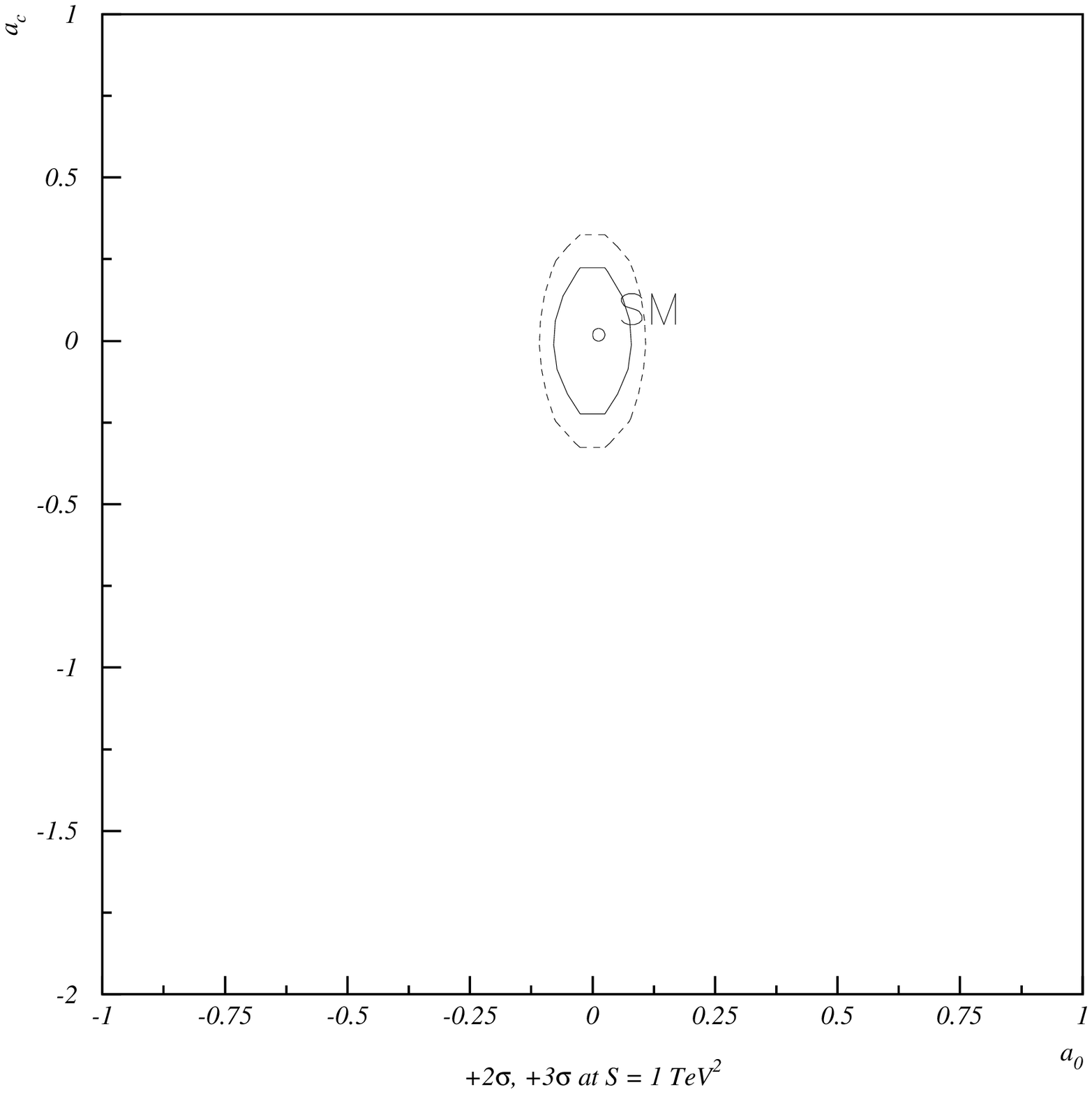}
\caption{Contour plots for $+2\sigma$, $+3\sigma$ deviations of
$\sigma(W^+W^-Z)$ at $\int {\cal L} = 100fb^{-1}, \sqrt{S}=1TeV$ }
\label{p18}
\end{minipage}\hfill
\end{figure}

\begin{figure}[h!]
\begin{minipage}[b]{.975\linewidth}
\centering
\includegraphics[width=\linewidth, height=3.8in, angle=0]{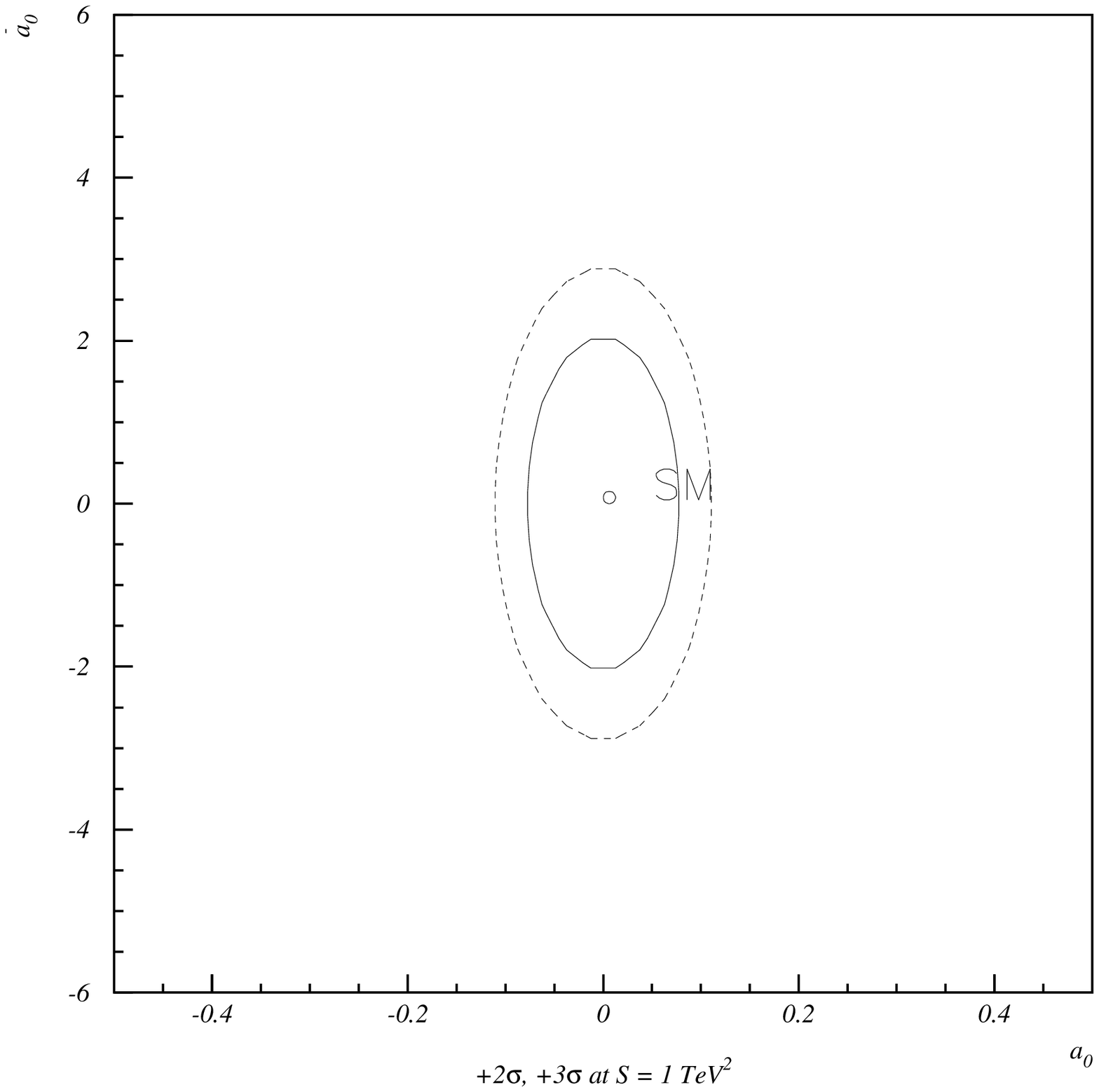}
\caption{ Contour plots for $+2\sigma$, $+3\sigma$ deviations of
$\sigma(W^+W^-Z)$ at $\int {\cal L} = 100fb^{-1}, \sqrt{S}=1TeV$ }
\label{p19}
\end{minipage}\hfill
\end{figure}

\begin{figure}[h!]
\begin{minipage}[b]{.975\linewidth} \centering
\includegraphics[width=\linewidth, height=3.8in, angle=0]{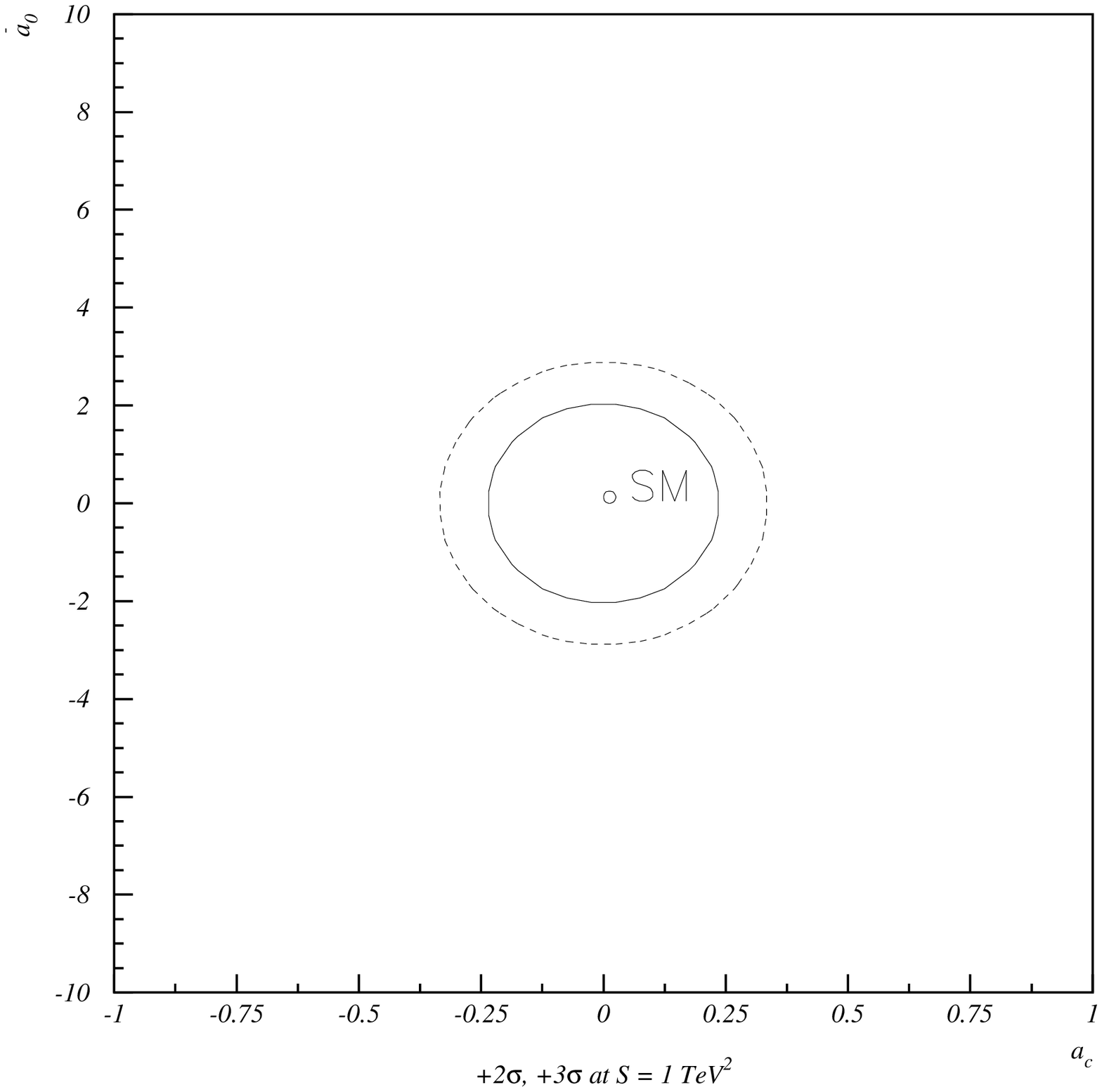}
\caption{Contour plots for $+2\sigma$, $+3\sigma$ deviations of
$\sigma(W^+W^-Z)$ at $\int {\cal L} = 100fb^{-1}, \sqrt{S}=1TeV$ }
\label{p20}
\end{minipage}\hfill
\end{figure}

\begin{figure}[h!]
\begin{minipage}[b]{.975\linewidth}
\centering
\includegraphics[width=\linewidth, height=3.8in, angle=0]{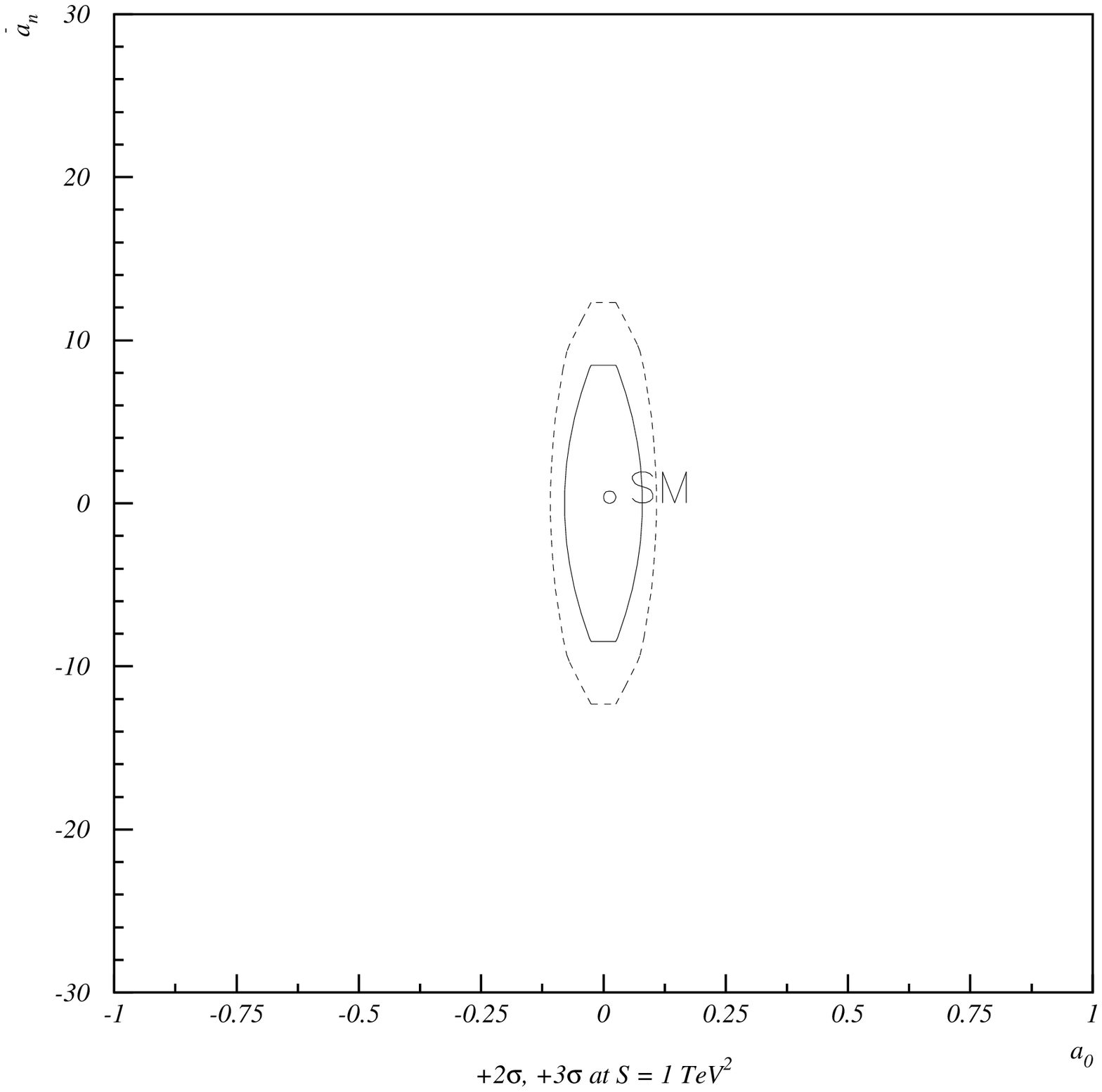}
\caption{ Contour plots for $+2\sigma$, $+3\sigma$ deviations of
$\sigma(W^+W^-Z)$ at $\int {\cal L} = 100fb^{-1}, \sqrt{S}=1TeV$ }
\label{p21}
\end{minipage}\hfill
\end{figure}

\begin{figure}[h!]
\begin{minipage}[b]{.975\linewidth} \centering
\includegraphics[width=\linewidth, height=3.8in, angle=0]{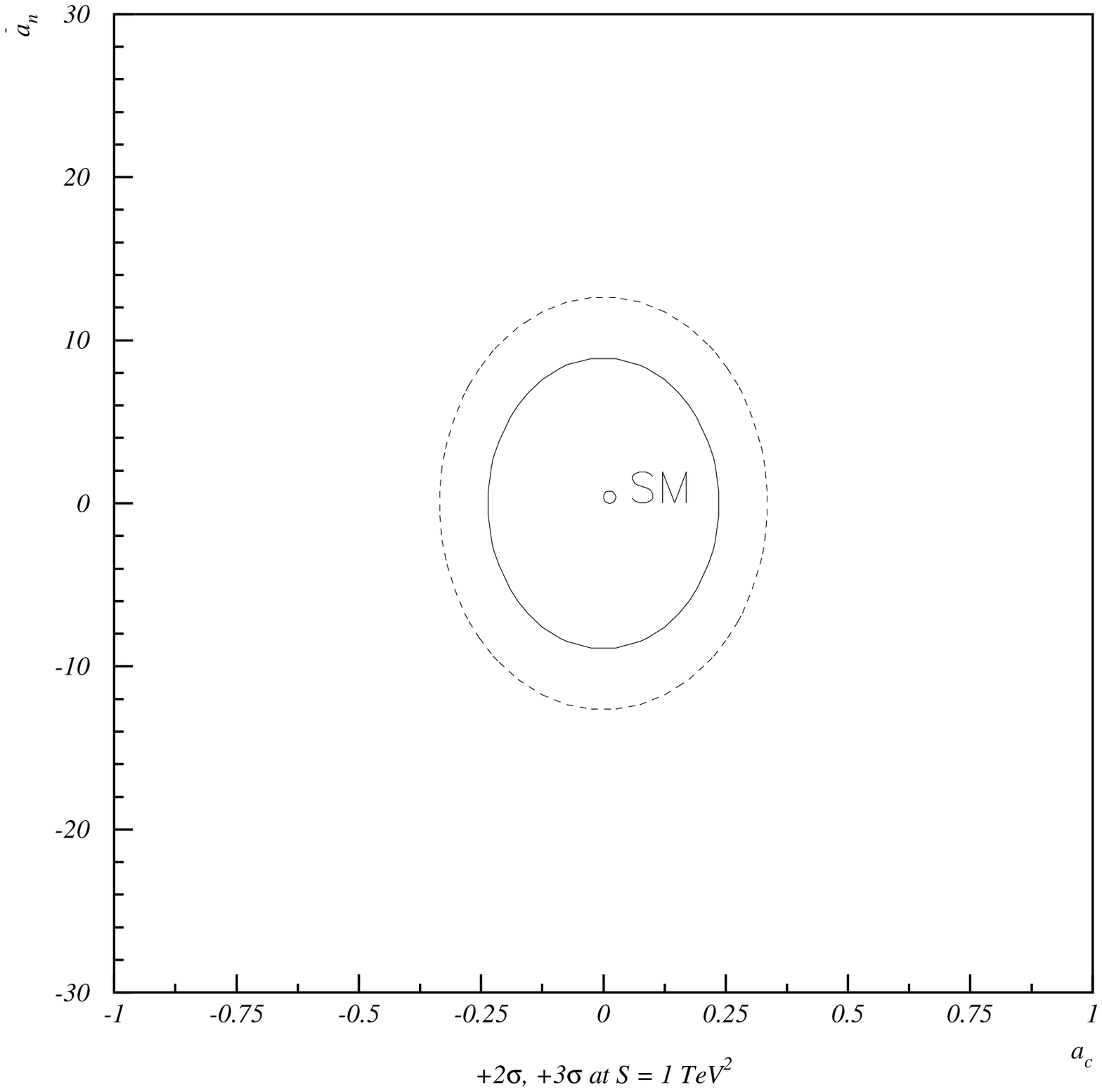}
\caption{Contour plots for $+2\sigma$, $+3\sigma$ deviations of
$\sigma(W^+W^-Z)$ at $\int {\cal L} = 100fb^{-1}, \sqrt{S}=1TeV$ }
\label{p22}
\end{minipage}\hfill
\end{figure}

The contour plots for different deviations from the SM
$\gamma\gamma\rightarrow W^+W^-$ and $\gamma\gamma\rightarrow W^+W^-Z$
total
cross sections at $\sqrt{S} = 1$ TeV, when two of the five
anomalous couplings are non-zero, are the most vivid and
interesting for investigation of
anomalous couplings. Figures \ref{p15} -- \ref{p22} presents
contour plots for different pairs of $a_i$. Here $\sigma$ means a variance of total cross section.

We have investigated the sensitivity of
$\gamma\gamma\rightarrow W^+W^-$ and
$\gamma\gamma\rightarrow W^+W^-Z$ to geniune
anomalous quartic couplins: $a_0$, $a_c$,
$\tilde{a}_0$, $a_n$, $\tilde{a}_n$ at the
centre-of-mass energy $\sqrt{S}=1$TeV.
We have assumed that all other couplings obtained
from $6$-dimensional operators and trilinear
couplings are eqaul to zero.
If the last are non-zero \cite{c6} one would
expect that limits obtained on the quartic
couplings can have been affected.

It is necessary to note that the sensitivities
of $W^+W^-$ and $W^+W^-Z$ productions in
$\gamma\gamma$ collisions are large in
comparison with these productions using $e^+e^-$
annihilation. It means that at high energies TESLA
has the great chance to reveal "new physics"
in the bosonic sector of the electroweak theory.

\end{document}